\documentclass[fleqn]{2020SCGE} 
\usepackage{siunitx}
\usepackage{hyperref}

\usepackage{booktabs}
\usepackage{lineno}
\usepackage{color}
\usepackage{verbatim}
\usepackage{float}

\usepackage{graphicx}
\usepackage{amsmath}
\usepackage{amssymb}	
\usepackage{diagbox}
\usepackage{threeparttable}

\usepackage{ulem}
\usepackage[numbers]{natbib}
\usepackage{longtable, threeparttablex, booktabs, url}
\usepackage{multirow} 
\usepackage{array} 
\usepackage{makecell} 
\usepackage{colortbl}
\usepackage{xcolor}
\usepackage{tabularx} 

\usepackage{color}

\begin{document}

\ensubject{subject}

\ArticleType{Article}
\SpecialTopic{SPECIAL TOPIC: }
\Year{2023}
\Month{xxx}
\Vol{xx}
\No{x}
\DOI{xx}
\ArtNo{000000}
\ReceiveDate{xx}
\AcceptDate{xx}
\AuthorMark{Yi Feng}

\AuthorCitation{ Yi Feng, Yong-Kun Zhang, Jintao Xie, Yuan-Pei Yang, Yuanhong Qu, Dengke Zhou, Di Li, Bing Zhang, Weiwei Zhu, Wenbin Lu, Jiaying Xu, Chenchen Miao, Shiyan Tian, Pei Wang, Ju-Mei Yao, Chen-Hui Niu, Jiarui Niu, Heng Xu, Jinchen Jiang, Dejiang Zhou, Zenan Liu, Chao-Wei Tsai, Zigao Dai, Xuefeng Wu, Fayin Wang, Jinlin Han, Kejia Lee, Renxin Xu, Yongfeng Huang, Yuanchuan Zou, Jinhuang Cao, Xianglei Chen, Jianghua Fang, Dongzi Li, Ye Li, Wanjin Lu, Jiawei Luo, Jintao Luo, Rui Luo, Fen Lyu, Bojun Wang, Weiyang Wang, Qin Wu, Mengyao Xue, Di Xiao, Wenfei Yu, Jianping Yuan, Chunfeng Zhang, Junshuo Zhang, Lei Zhang, Songbo Zhang, Rushuang Zhao, Yuhao Zhu}
\title{Multi-year Polarimetric Monitoring of Four CHIME-Discovered Repeating Fast Radio Bursts with FAST}

\author[1,2]{Yi Feng\footnote{Corresponding author. Email: yifeng@zhejianglab.org}}{}
\author[3]{Yong-Kun Zhang}{}
\author[4]{Jintao Xie}{}
\author[5,6]{Yuan-Pei Yang}{}
\author[7,8]{Yuanhong Qu}{}
\author[1]{Dengke Zhou}{}
\author[9,3]{Di Li\footnote{Corresponding author. Email: dili@mail.tsinghua.edu.cn}}{}
\author[7,8]{\\Bing Zhang\footnote{Corresponding author. Email: bing.zhang@unlv.edu}}{}
\author[3,10]{Weiwei Zhu}{}
\author[11]{Wenbin Lu}{}
\author[1]{Jiaying Xu}{}
\author[1]{Chenchen Miao}{}
\author[12]{Shiyan Tian}{}
\author[3,10]{Pei Wang}{}
\author[13]{\\Ju-Mei Yao}{}
\author[14]{Chen-Hui Niu}{}
\author[3]{Jiarui Niu}{}
\author[3]{Heng Xu}{}
\author[3]{Jinchen Jiang}{}
\author[3]{Dejiang Zhou}{}
\author[15]{Zenan Liu}{}
\author[3,10,16]{\\Chao-Wei Tsai}{}
\author[17]{Zigao Dai}{}
\author[6]{Xuefeng Wu}{}
\author[15]{Fayin Wang}{}
\author[3]{Jinlin Han}{}
\author[18,19,3]{Kejia Lee}{}
\author[18,19]{Renxin Xu}{}
\author[15, 20]{\\Yongfeng Huang}{}
\author[12]{Yuanchuan Zou}{}
\author[3]{Jinhuang Cao}{}
\author[3]{Xianglei Chen}{}
\author[1]{Jianhua Fang}{}
\author[21]{Dongzi Li}{}
\author[6]{Ye Li}{}
\author[3]{\\Wanjin Lu}{}
\author[22]{Jiawei Luo}{}
\author[23]{Jintao Luo}{}
\author[24]{Rui Luo}{}
\author[25]{Fen Lyu}{}
\author[3]{Bojun Wang}{}
\author[16,18,19]{Weiyang Wang}{}
\author[15]{Qin Wu}{}
\author[3]{\\Mengyao Xue}{}
\author[6]{Di Xiao}{}
\author[26]{Wenfei Yu}{}
\author[13]{Jianping Yuan}{}
\author[3]{Chunfeng Zhang}{}
\author[3, 16]{Junshuo Zhang}{}
\author[3]{Lei Zhang}{}
\author[6]{\\Songbo Zhang}{}
\author[27]{Rushuang Zhao}{}
\author[3]{Yuhao Zhu}{}

\address[1]{Research Center for Astronomical Computing, Zhejiang Laboratory, Hangzhou {\rm 311100}, China}
\address[2]{Institute for Astronomy, School of Physics, Zhejiang University, Hangzhou {\rm 310027}, China}
\address[3]{National Astronomical Observatories, Chinese Academy of Sciences, Beijing {\rm 100101}, China}
\address[4]{School of Computer Science and Engineering, Sichuan University of Science and Engineering, Yibin {\rm 644000}, China}
\address[5]{South-Western Institute for Astronomy Research, Yunnan University, Kunming {\rm 650504}, China}
\address[6]{Purple Mountain Observatory, Chinese Academy of Sciences, Nanjing {\rm 210023}, China}
\address[7]{Nevada Center for Astrophysics, University of Nevada, Las Vegas, NV {\rm 89154}, USA}
\address[8]{Department of Physics and Astronomy, University of Nevada, Las Vegas, NV {\rm 89154}, USA}
\address[9]{New Cornerstone Science Laboratory, Department of Astronomy, Tsinghua University, Beijing {\rm 100084}, China}
\address[10]{Institute for Frontiers in Astronomy and Astrophysics, Beijing Normal University, Beijing {\rm 102206}, China}
\address[11]{Departments of Astronomy and Theoretical Astrophysics Center, UC Berkeley, Berkeley, CA {\rm 94720}, USA}
\address[12]{Department of Astronomy, School of Physics, Huazhong University of Science and Technology, Wuhan {\rm 430074}, China}
\address[13]{Xinjiang Astronomical Observatory, Chinese Academy of Sciences, Urumqi, Xinjiang {\rm 830011}, China}
\address[14]{Institute of Astrophysics, Central China Normal University, Wuhan {\rm 430079}, Hubei, China}
\address[15]{School of Astronomy and Space Science, Nanjing University, Nanjing {\rm 210023}, China}
\address[16]{University of Chinese Academy of Sciences, Beijing {\rm 100049}, China}
\address[17]{Department of Astronomy, School of Physical Sciences, University of Science and Technology of China, Hefei {\rm 230026}, China}
\address[18]{Department of Astronomy, Peking University, Beijing {\rm 100871}, China}
\address[19]{Kavli Institute for Astronomy and Astrophysics, Peking University, Beijing {\rm 100871}, China}
\address[20]{Key Laboratory of Modern Astronomy and Astrophysics (Nanjing University), Ministry of Education, Nanjing {\rm 210023}, China}
\address[21]{Department of Astrophysical Sciences, Princeton University, Princeton, NJ {\rm 08544}, USA}
\address[22]{College of Physics, Hebei Normal University, Shijiazhuang {\rm 050024}, China}
\address[23]{National Time Service Center, Chinese Academy of Sciences, Xi’an, {\rm 710600}, China}
\address[24]{Department of Astronomy, School of Physics and Materials Science, Guangzhou University, Guangzhou {\rm 510006}, China}
\address[25]{Institute of Astronomy and Astrophysics, Anqing Normal University, Anqing {\rm 246133}, China}
\address[26]{Shanghai Astronomical Observatory, Chinese Academy of Sciences, Shanghai {\rm200030}, China}
\address[27]{Guizhou Provincial Key Laboratory of Radio Astronomy and Data Processing, Guizhou Normal University, Guiyang {\rm 550001}, China}

\abstract{
Fast radio bursts (FRBs) are bright, millisecond-duration radio emissions originating from cosmological distances. 
In this study, we report multi-year polarization measurements of four repeating FRBs initially discovered by the Canadian Hydrogen Intensity Mapping Experiment (CHIME): FRBs~20190117A, 20190208A, 20190303A, and 20190417A. We observed the four repeating FRBs with the Five-hundred-meter Aperture Spherical Radio Telescope (FAST), detecting a total of 66 bursts.
Two bursts from FRB~20190417A exhibit a circular polarization signal-to-noise ratio greater than 7, with the highest circular polarization fraction recorded at 35.7\%. While the bursts from FRBs 20190208A and 20190303A are highly linearly polarized, those from FRBs~20190117A and 20190417A show depolarization due to multi-path propagation, with $\sigma_{\mathrm{RM}} = 2.78 \pm 0.05 \, \text{rad m}^{-2}$ and $5.19 \pm 0.09 \, \text{rad m}^{-2}$, respectively.
The linear polarization distributions among five repeating FRBs—FRBs~20190208A, 20190303A, 20201124A, 20220912A, and 20240114A—are nearly identical but show distinct differences from those of non-repeating FRBs. FRBs~20190117A, 20190303A, and 20190417A exhibit substantial rotation measure (RM) variations between bursts, joining other repeating FRBs in this behavior. Combining these findings with published results, 64\% of repeating FRBs show RM variations greater than 50\,rad m$^{-2}$, and 21\% exhibit RM reversals. A significant proportion of repeating FRBs reside in a dynamic magneto-ionic environment.
The structure function of RM variations shows a power-law index of $\gamma \sim (0-0.8)$, corresponding to a shallow power spectrum $\alpha = -(\gamma + 2) \sim -(2.0-2.8)$ of turbulence, if the RM variations are attributed to turbulence. This suggests that the variations are dominated by small-scale RM density fluctuations. We perform K-S tests comparing the RMs of repeating and non-repeating FRBs, which reveal a marginal dichotomy in the distribution of their RMs. We caution that the observed dichotomy may be due to the small sample size and selection biases.}



\maketitle


\begin{multicols}{2}
\section{Introduction} \label{sec:intro}
Fast radio bursts (FRBs) are short (lasting from microseconds to milliseconds), bright radio transients first discovered by Ref.~\cite{2007Sci...318..777L}. Their progenitors and radiation mechanisms remain unknown \citep{zhangreview2023}. Repeating FRBs form a subset of FRBs that emit multiple bursts from the same source over time. Out of approximately 800 known FRBs, around 60 are classified as repeating FRBs\footnote{https://blinkverse.zero2x.org} \citep{blinkverse}. The repetitive nature of these bursts enables astronomers to investigate various properties of each burst—such as their spectral, temporal, and polarimetric characteristics—providing valuable insights into their potential origins, radiation mechanisms, and surrounding environments.

Polarization is a fundamental property of FRBs, carrying crucial information about their intrinsic characteristics and the environments in which they occur. Precise measurements of polarization properties form the foundation for studying the surrounding environment, radiation mechanisms, and classification of FRBs. These measurements provide key evidence for unraveling the physical origins, radiation mechanisms, and environmental features of these bursts.
The Faraday effect, in particular, plays a significant role by causing the polarization plane of electromagnetic waves to rotate. This rotation is quantified by the rotation measure (RM), which encodes information about the medium the FRB traverses, such as magnetic field strength and electron density. For instance, observations of FRB~20121102A have revealed a remarkably high and time-varying RM, interpreted as evidence that this FRB resides in a dynamic, strongly magnetized, ionized environment \citep{121102rm, 2021ApJ...908L..10H, 2023ATel15980....1F}. Possible environments include expanding supernova remnants, pulsar wind nebulae, massive binary systems, and so on \citep{2017ApJ...847...22Y, 2018ApJ...861..150P, 2018ApJ...868L...4M, 2021ApJ...923L..17Z, 2022NatCo..13.4382W, 2022MNRAS.510L..42K, 2022ApJ...928L..16Y, 2023ApJ...942..102Z, 2023MNRAS.520.2039Y}.
The RM of FRB~20190520B underwent rapid changes, increasing from approximately 3000\,rad\,m\(^{-2}\) in March 2021 to about 10,000\,rad\,m\(^{-2}\) in June 2021. It then reversed to approximately \(-24,000\)\,rad\,m\(^{-2}\) over the next five months \citep{reshma23}. This first-of-its-kind RM reversal strongly suggests that FRB~20190520B is located in a binary system, potentially accompanied by a massive star \citep{reshma23}.
Similarly, the highly variable RM observed in FRB~20201124A revealed a dynamic environment close to the FRB’s central engine, at a distance scale of about 1\,AU \citep{2022Natur.609..685X}. Oscillation structures as functions of frequency in the polarization profiles indicate the presence of a strong magnetic field around the source \citep{2022Natur.609..685X}. Additionally, prominent levels of circular polarization were observed in a substantial number of bursts from this active repeating FRB source, further supporting a magnetospheric origin for FRB emission \citep{Wang2022}.

\begin{table*}
\centering
\caption{Observation positions of 4 FRBs. Column (1): FRB name; Column (2, 3, 4): FAST observation coordinate and uncertainty; Column (5, 6, 7, 8): CHIME estimated position and uncertainty; Column (9): Angular distance between the positions measured by FAST and CHIME .}
\label{tab:position}
    \begin{tabular}[t]{c|ccc|cccc|c}
    \hline
        \multirow{2}{*}{Source}  & \multicolumn{3}{c|}{FAST} & \multicolumn{4}{c|}{CHIME} & \multirow{2}{*}{Dist.('')} \\
        \cline{2-8}
                & RA  & Dec        & $\sigma$('') & RA & $\sigma_{\rm{RA}}$('') & Dec & $\sigma_{\rm{Dec}}$('')  \\
        \hline
        FRB~20190117A & 22h06m43s  &  +17d20m28s  & 156 & 22h06m38s & 13 & 17d22m06s & 13 &  121.4\\ 
        FRB~20190208A & 18h54m17s  &  +46d55m27s  & 156 & 18h54m07s & 12 & 46d55m20s & 13 &  102.6\\ 
        FRB~20190303A & 13h51m58s  &  +48d07m20s  & 156 & 13h51m59s & 11 & 48d07m16s & 12 &   10.8\\ 
        FRB~20190417A & 19h39m22s  &  +59d18m58s  & 156 & 19h39m04s & 15 & 59d19m55s & 16 &  149.1\\ 
    \hline
    \end{tabular}
\end{table*}

The polarization position angle (PA) and degrees of linear and circular polarization can be used to trace the intrinsic radiation mechanisms and propagation processes \citep{Qu&Zhang2023}. 
The models inside the magnetar magneotosphere invoking charged bunches via curvature radiation and inverse Compton scattering can mainly produce high linear polarization and non-negligible circular polarization in different viewing angles \citep{Wang2022,Liu2023cp,Qu&Zhang2023,Qu&Zhang2024}.
The relativistic shock models outside the magnetosphere mainly produce linear polarized X-mode waves \citep{Plotnikov,Sironi2021}.
Observationally, the burst with a 90\% circular polarization degree and orthogonal modes in linear polarization detected in FRB~20201124A strongly constrain the FRB radiation mechanisms, among which the pulsar-like magnetospheric models are more plausible \citep{2024NSRev..12..293J}.
The PA of FRB~20180301A showed various short-timescale swings, which are hypothesized to originate within the magnetosphere of a magnetar \citep{luo2020}. The ``S"-shaped PA swing also suggests a magnetospheric origin \citep{Mckinven2024}. FRB~20201124A exhibited a sudden PA jump within about 1\,ms. This jump could be due to the superposition of two orthogonal emission modes, and the millisecond timescale of the jump suggests that the FRB emission originates from the complex magnetosphere of a magnetar \citep{Niu_2024}.

Recently, long-term systematic measurements of the polarization of repeating fast radio bursts have been released, and an overall picture of the polarization properties of repeating FRBs is beginning to emerge. For example, Ref. \citep{chime_repeaterRM} presents a multiyear polarimetric study of 12 repeating FRBs using the CHIME telescope. They observe significant RM variations from many sources in their sample, including RM changes of several hundred rad/m$^{2}$ over month timescales from FRBs~20181119A, 20190303A, and 20190417A, as well as more modest RM variability (RM of a few tens of rad/m$^{2}$) from FRBs~20181030A, 20190208A, 20190213B, and 20190117A over equivalent timescales. Repeating FRBs occupy more dynamic magneto-ionic environments compared to pulsars in the Milky Way and the Magellanic Clouds. Ref. \citep{cherry2025} presents a polarimetric study of 28 repeating FRBs observed by the CHIME telescope, analyzing 75 bursts (63 with significant RM measurements) between 2019 and 2024. They observe temporal RM variations in practically all repeating FRBs. Following the trend revealed by Ref. \citep{chime_repeaterRM}, Ref. \citep{cherry2025} further categorizes the repeating FRBs into two distinct groups: those inhabiting dynamic RM environments and those in stable magneto-ionic media, based on $\sigma$(RM)/$|$RM$|$. RM sign changes were observed in FRBs~20190303A, 20200929C, and possibly 20180916B, joining FRB~20190520B \citep{niu22} and FRB~20180301A \citep{2023MNRAS.526.3652K}, which indicate magnetic field reversals. The polarization properties of repeating FRBs have also been compared with those of non-repeating FRBs. Ref. \citep{cherry2025} shows a marginal dichotomy in the distribution of electron-density-weighted parallel-component line-of-sight magnetic fields between repeating and non-repeating FRBs.

In this paper, we report on multi-year polarimetric monitoring of four repeating FRBs discovered by the Canadian Hydrogen Intensity Mapping Experiment (CHIME), namely FRBs~20190117A, 20190208A, 20190303A, and 20190417A, using the Five-hundred-meter Aperture Spherical Radio Telescope (FAST) \citep{2011IJMPD..20..989N, li16, 2018IMMag..19..112L}. The paper is structured as follows: Section~\ref{sec:obs} describes our observations and data processing procedures; the results are presented in Section~\ref{sec:re}, followed by a discussion in Section~\ref{sec:dis}. Finally, Section~\ref{sec:con} summarizes our conclusions.

\section{Observations and Data Reduction} \label{sec:obs}

We observed four repeating FRBs~20190117A, 20190208A, 20190303A, and 20190417A discovered by CHIME with FAST from 2020 to 2023. As the positions of these repeaters are reported with error bars of 10-20 arcminutes, much larger than the beam size at FAST, we first needed to determine their positions within the FAST beam with all beams of the 19-beam receiver \citep{2019SCPMA..6259502J}.    
For the first observations of FRB~20190117A and FRB~20190208A, the observations were conducted using all beams of the 19-beam receiver in snapshotdec or onoff mode\footnote{A description of the two observation modes can be found at https://fast.bao.ac.cn/cms/article/24/.}. For FRB~20190117A, we detected one burst in the central beam on 5 December 2021. Using this burst, the estimated position of FRB~20190117A is (Right Ascension--RA, Declination--Dec) = (22$^h$06$^m$43$^s$, +17$^\circ$20$'$28$''$) (J2000) with an error circle of 2.6$'$. Similarly, using one burst detected from FRB~20190208A on 6 July 2021, the estimated position of FRB~20190208A is (RA, DEC) = (18$^h$54$^m$17$^s$, +46$^\circ$55$'$27$''$) (J2000) with an error circle of 2.6$'$. For FRB~20190303A and FRB~20190417A, we used the positions reported in Ref.~\cite{feng22}, i.e., (RA, DEC) = (13$^h$51$^m$58$^s$, +48$^\circ$07$'$20$''$) (J2000) for FRB~20190303A, and (RA, DEC) = (19$^h$39$^m$22$^s$, +59$^\circ$18$'$58$''$) (J2000) for FRB~20190417A. Subarcminute localization of these four FRBs was later achieved with CHIME baseband data \citep{2023ApJ...950..134M}. The positions determined from CHIME baseband data and FAST are listed in Table~\ref{tab:position}. We note that the positions determined with FAST are consistent with the CHIME subarcminute localization. Such off-axis illumination at FAST does not affect the polarization measurements \citep{fastbeam, luo2020}. 

The successive observations were conducted using the central beam of the 19-beam receiver pointed at the estimated positions. The 19-beam receiver, with the frequency range between 1000 and 1500\,MHz, provides two data streams (one for each hand of linear polarization). The data streams are processed with the Reconfigurable Open Architecture Computing Hardware–version 2 (ROACH2) signal processor \citep{2019SCPMA..6259502J}. The output full-Stokes data files are recorded as 8 bit-sampled search mode PSRFITS \citep{2004PASA...21..302H} files every 49.152 $\mu$s with 4096 frequency channels. FRBs~20190117A, 20190208A, 20190303A were observed approximately once a month. FRB~20190417A was observed more frequently to investigate its possible periodicity (Zhang et al., in prep). The typical observation duration is half an hour. The dates, durations, and modes of all observations are listed in Table~\ref{tab:obsLog}. 

\renewcommand{\arraystretch}{1.1}
\begin{table*}[htbp]
\centering
\setlength{\tabcolsep}{9pt}
\caption{A table summarizing the observations. Column (1): Observation date; Column (2): Observation duration in hours; Column (3): Observation mode; Column (4): Number of bursts detected.}
\label{tab:obsLog}
\begin{tabular}{cccc|cccc}
\hline
Obs. Date & Obs. Length (hour)  & Obs. Mode & $N_{\rm{burst}}$  & Obs. Date & Obs. Length  (hour) & Obs. Mode & $N_{\rm{burst}}$ \\
\hline
\multicolumn{8}{c}{{FRB~20190117A}} \\
\hline
20211120 & 2   & snapshotdec & 0 & 20220526 & 0.5  & tracking    & 0 \\
20211205 & 2   & snapshotdec & 1 & 20220624 & 0.5  & tracking    & 0 \\
20211214 & 2   & tracking    & 0 & 20220721 & 0.5  & tracking    & 0 \\
20220205 & 1   & tracking    & 0 & 20220801 & 0.5  & tracking    & 0 \\
20220223 & 0.5 & tracking    & 0 & 20220907 & 0.5  & tracking    & 0 \\
20220330 & 0.5 & tracking    & 0 & 20221004 & 0.5  & tracking    & 0 \\
20220424 & 0.5 & tracking    & 0 & 20221021 & 0.5  & tracking    & 0 \\
\hline
\multicolumn{8}{c}{{FRB~20190208A}} \\
\hline
20210630 & 1   & onoff       & 0 & 20220527 & 0.5  & tracking    & 0 \\
20210706 & 1   & onoff       & 1 & 20220624 & 0.5  & tracking    & 0 \\
20210712 & 2   & tracking    & 4 & 20220801 & 0.5  & tracking    & 0 \\
20211016 & 2   & tracking    & 0 & 20220909 & 0.5  & tracking    & 0 \\
20220223 & 0.5 & tracking    & 0 & 20221008 & 0.5  & tracking    & 1 \\
20220401 & 0.5 & tracking    & 0 & 20221024 & 0.5  & tracking    & 0 \\
20220424 & 0.5 & tracking    & 0 &          &      &             &   \\
\hline
\multicolumn{8}{c}{{FRB~20190303A}} \\
\hline
20211019 & 2   & tracking    & 0 & 20220624 & 0.5  & tracking    & 0 \\
20220222 & 0.5 & tracking    & 4 & 20220801 & 0.5  & tracking    & 1 \\
20220330 & 0.5 & tracking    & 0 & 20220907 & 0.5  & tracking    & 4 \\
20220424 & 0.5 & tracking    & 0 & 20220924 & 0.5  & tracking    & 2 \\
20220529 & 0.5 & tracking    & 0 & 20221022 & 0.5  & tracking    & 1 \\
20220603 & 0.5 & tracking    & 0 & 20231020 & 1    & tracking    & 0 \\
20220607 & 0.5 & tracking    & 1 &          &      &             &   \\
\hline
\multicolumn{8}{c}{{FRB~20190417A}} \\
\hline
20200825 & 1   & tracking    & 0 & 20220516 & 0.5  & tracking    & 4 \\
20200828 & 1   & tracking    & 0 & 20220519 & 0.5  & tracking    & 0 \\
20200904 & 1   & tracking    & 0 & 20220524 & 0.33 & tracking    & 0 \\
20201114 & 1   & tracking    & 0 & 20220605 & 0.5  & tracking    & 2 \\
20210325 & 1   & tracking    & 0 & 20220609 & 0.5  & tracking    & 0 \\
20211003 & 2   & tracking    & 9 & 20220614 & 0.5  & tracking    & 1 \\
20220228 & 0.5 & tracking    & 7 & 20220622 & 0.5  & tracking    & 1 \\
20220309 & 0.5 & tracking    & 1 & 20220628 & 0.5  & tracking    & 0 \\
20220312 & 0.5 & tracking    & 1 & 20220710 & 0.5  & tracking    & 0 \\
20220313 & 0.5 & tracking    & 0 & 20220713 & 0.5  & tracking    & 4 \\
20220315 & 0.5 & tracking    & 0 & 20220729 & 0.5  & tracking    & 2 \\
20220326 & 0.5 & tracking    & 0 & 20220801 & 0.5  & tracking    & 5 \\
20220331 & 0.5 & tracking    & 0 & 20220808 & 0.5  & tracking    & 2 \\
20220414 & 0.5 & tracking    & 0 & 20220813 & 0.5  & tracking    & 2 \\
20220426 & 0.5 & tracking    & 0 & 20220904 & 0.5  & tracking    & 0 \\
20220509 & 0.5 & tracking    & 6 & 20221013 & 0.5  & tracking    & 0 \\
20220515 & 0.5 & tracking    & 0 & 20221123 & 0.5  & tracking    & 0 \\
\hline
\end{tabular}
\end{table*}

We used \textsc{DRAFTS}\footnote{\url{https://github.com/SukiYume/DRAFTS}} for burst searching, a deep learning-based tool for detection of radio transients \citep{2025ApJS..276...20Z}. When searching the data from FAST observations of the four FRBs, we first de-dispersed the FAST data using the fiducial dispersion measure (DM) values corresponding to the repeating FRBs: ${\rm DM} = 397 \,\rm pc\,cm^{-3}$ for FRB~20190117A, ${\rm DM} = 580 \,\rm pc\,cm^{-3}$ for FRB~20190208A, ${\rm DM} = 222 \,\rm pc\,cm^{-3}$ for FRB~20190303A, and ${\rm DM} = 1378 \,\rm pc\,cm^{-3}$ for FRB~20190417A. Subsequently, we unified the time resolution of the data to $196.608\,\rm \mu s$ (down-sampled by a factor of 4 from the typical time resolution of $49.152\,\rm \mu s$ for FAST) to enhance the signal-to-noise ratio (S/N).

After down-sampling, we divided the data into segments of 512 time samples, resizing each segment to $512 \times 512$ pixels, and inputting them into the classification model in \textsc{DRAFTS} for inference. The model provided the probability of an FRB burst occurring within each segment. We selected the segments where the predicted probability exceeded 50\% as potential burst candidates, excluding those with low S/N. Finally, we performed manual post-processing to determine the arrival time of the bursts based on their positions within the data segments.

To remove radio frequency interference (RFI), we used the PSRCHIVE software package \citep{2004PASA...21..302H}. A median filter was applied to each burst in the frequency domain using the paz command, and RFI was manually mitigated in each burst using the pazi command. 
Polarization calibration was performed by correcting for differential gain and phase between the receptors. This was achieved through separate measurements of a noise diode signal injected at a $45^{\circ}$ angle relative to the linear receptors, using the single-axis model with the PSRCHIVE software package.
We measured the RM using the Stokes QU-fitting method \citep{2012MNRAS.421.3300O}, selecting positions where the signal-to-noise ratio exceeded 3 for RM fitting. We corrected the linear polarization by derotating it using the measured RM.

We calculated the degrees of linear polarization and circular polarization for bursts with RM detection. The measured linear polarization is overestimated in the presence of noise. Therefore we use the frequency-averaged, de-biased total linear polarization \citep{2001ApJ...553..341E, askap20}\footnote{A typographical error in Ref.~\cite{2001ApJ...553..341E} was corrected in Ref.~\cite{askap20}.} :
\begin{equation} \label{eq:L_de-bias}
    L_{{\mathrm{de\mbox{-}bias}}} =
    \begin{cases}
      \sigma_I \sqrt{\left(\frac{L_{i}}{\sigma_I}\right)^2 - 1} & \text{if $\frac{L_{i}}{\sigma_I} > 1.57$} \\
      0 & \text{otherwise} ,
    \end{cases}
\end{equation}
where $\sigma_I$ is the Stokes I off-pulse standard deviation and $L_i$ is the measured frequency-averaged linear polarization of time sample $i$. 
We defined the degree of linear polarization 
as ($\Sigma_{i} L_{\mathrm{de\mbox{-}bias},i}$)/($\Sigma_{i}I_i$)
and that of circular polarization as ($\Sigma_{i} V_i$)/($\Sigma_{i}I_i$), where the summation is over the time samples in one burst and $V_i$ is the measured frequency-averaged circular polarization of time sample $i$. 
Defining $I = \Sigma_{i}I_i$, $L = \Sigma_{i} L_{\mathrm{de\mbox{-}bias},i}$ and $V = \Sigma_{i}V_i$, uncertainties on the linear polarization fraction and circular polarization fraction are calculated as:
\begin{equation} \label{eq:uncertainty}
    \sigma_{\rho/I} = \frac{\sqrt{N+N\frac{\rho^2}{I^2}}}{I}\sigma_{I},
\end{equation}
where $N$ is the number of time samples of the burst (signal exceeds the noise by 3$\sigma$), and $\rho = L,V$ for linear and circular polarization fraction, respectively.

We then determined the frequency range of each burst. A detailed description of the method can be found in Ref.~\cite{xie2024FRB20240114A}, while a brief introduction is provided here. The frequency range was calculated based on the cumulative distribution function (CDF) of the burst spectrum, as described in Ref.~\cite{xie2024FRB20240114A}. The CDF was normalized using smoothing filters and the Asymmetrically Reweighted Penalized Least Squares algorithm. The frequency indices marking the start and end points of the burst range were identified from the first derivative of the smoothed CDF. The lower and upper boundaries were determined based on the positivity of the first derivative relative to the CDF threshold of 0.5, with uncertainties assessed using the bootstrap method.

\section{Results} \label{sec:re}
We detected 1, 5, 13, 47 bursts for FRBs~20190117A, 20190208A, 20190303A, and 20190417A, respectively, with a total of 66 bursts. The numbers of detections are listed in Table~\ref{tab:obsLog}. All the polarization pulse profiles and dynamic spectra are included in Supplementary Materials. The highest burst rates are 0.5, 2, 8, and 14 $\mathrm{bursts~hr^{-1}}$ for FRBs~20190117A, 20190208A, 20190303A, and 20190417A, respectively. The mean burst rates are 0.08, 0.48, 1.53, and 2.26 $\mathrm{bursts~hr^{-1}}$ for the same FRBs. While the mean burst rates were calculated over the entire exposure period, the highest burst rates were calculated based on each individual observation session. Notably, FRB~20190417A exhibits the highest burst rate among the four repeaters. This may be a selection effect, as we observed FRB~20190417A during an active period, allowing us to detect more bursts. The average burst rates for these four repeaters at CHIME are on the order of 0.1 bursts per hour, much lower than those at FAST. The fluence completeness limits at CHIME are a few Jy ms, much larger than the $\sim$0.02 Jy ms at FAST. This difference in burst rates could therefore be due to variations in sensitivity and burst rates at different frequencies. Although the burst rates are higher than those measured at CHIME, they are still much lower than those of FRBs~20121102A, 20201124A, 20220912A, 20240114A. For example, FRBs~20121102A, 20201124A, 20220912A, and 20240114A had peak burst rates of 122 hr$^{-1}$ \citep{li21}, 542 hr$^{-1}$ \citep{2022RAA....22l4002Z}, 390 hr$^{-1}$ \citep{zhang2023}, and $\sim$500 hr$^{-1}$ \citep{2024ATel16505....1Z}, respectively, all recorded at FAST. Even a (sub)-100-meter radio telescope recorded a burst rate of 260 $\mathrm{hr^{-1}}$ for FRB~20240114A \citep{xie2024FRB20240114A}. Therefore, our sample does not represent the most active repeaters. The times of arrival, RM, degrees of linear and circular polarization, and the lower and upper frequencies of the 66 bursts are listed in Table~\ref{tab:burst}.

Bursts 13 and 32 of FRB~20190417A exhibit circular polarization with signal-to-noise ratios greater than 7. The circular polarization fractions are $11.9 \pm 1.0$\% and $35.7 \pm 3.7$\%, respectively. Circular polarization is a relatively rare phenomenon among repeating FRBs. Although there are over 60 repeating FRB sources \citep{blinkverse}, circular polarization has been observed in only five of them: FRBs~20121102A, 20190520B \citep{2022SciBu..67.2398F}, 20201124A \citep{2022Natur.609..685X}, 20220912A \citep{zhang2023,feng2024}, and 20240114A \citep{xie2024FRB20240114A}. The largest circular polarization fraction observed in our sample is $35.7 \pm 3.7$\% for FRB~20190417A, which is smaller than the typical $\sim$60\% to $\sim$90\% seen in other repeating FRBs. For example, the largest circular polarization fraction is about 64\% for FRB~20121102A \citep{2022SciBu..67.2398F}, about 70\% for FRB~20220912A \citep{zhang2023}, and about 65\% for FRB~20240114A \citep{xie2024FRB20240114A}. The highest recorded circular polarization fraction is approximately 90\% for FRB~20201124A \citep{2024NSRev..12..293J}. However, we note that the high polarization fractions observed in other repeaters are rare outliers, typically occurring in about 1 in 1000 bursts. Since FRB~20190417A has only two bursts with detected circular polarization, it is difficult to make a direct comparison to these outliers. To make a proper comparison, more bursts from FRB~20190417A will be needed.                                            

\end{multicols}
\renewcommand{\arraystretch}{1.1}
\begingroup
\setlength{\LTcapwidth}{0.9\textwidth} 
\begin{longtable}{cccccccc}
\caption{ Measured properties of 4 FRBs. Column (1): FRB name; Column (2): Burst index;  Column (3): Modified Julian dates referenced to infinite frequency at the Solar System barycentre; Column (4): RM obtained by Stokes QU-fitting; Column (5): Degree of linear polarization; Column (6): Degree of circular polarization; Column (7): Lower frequency of bursts; Column (8): Upper frequency of bursts. It is important to note that the lower boundary near 1000\,MHz and the upper boundary around 1500\,MHz are due to the proximity to the limits of the observational band.}
\label{tab:burst}\\\toprule
\endfirsthead
\caption{Continued:}\\\toprule
\endhead
\endfoot
\bottomrule
\endlastfoot
Source Name & Burst ID & MJD & RM  (rad~m$^{-2}$)& \% Linear & \% Circular  & Freq$_{\rm lower}$ (MHz)  & Freq$_{\rm upper}$ (MHz) \\
\midrule
\midrule
FRB 20190117A & 1 & 59553.40588624 & $-165_{-17}^{+20}$ & $28.9_{-2.7}^{+2.7}$ & $4.7_{-2.6}^{+2.6}$ & $1000_{-1}^{+1}$ & $1120_{-7}^{+7}$\\ 
\midrule
FRB 20190208A & 1 & 59407.62754923 & $29_{-7}^{+8}$ & $100.2_{-2.9}^{+2.9}$ & $-7.5_{-2.0}^{+2.0}$ & $1000_{-1}^{+1}$ & $1115_{-3}^{+3}$\\ 
FRB 20190208A & 2 & 59407.65927024 & $14_{-4}^{+6}$ & $87.6_{-8.4}^{+8.4}$ & $-16.7_{-6.4}^{+6.4}$ & $1047_{-2}^{+2}$ & $1279_{-24}^{+24}$\\ 
FRB 20190208A & 3 & 59407.65926490 & $21_{-8}^{+8}$ & $81.5_{-7.6}^{+7.6}$ & $-5.9_{-5.9}^{+5.9}$ & $1000_{-1}^{+1}$ & $1124_{-2}^{+2}$\\ 
FRB 20190208A & 4 & 59407.65926479 & $17_{-4}^{+6}$ & $100.6_{-2.5}^{+2.5}$ & $-9.1_{-1.8}^{+1.8}$ & $1000_{-1}^{+1}$ & $1136_{-2}^{+2}$\\ 
FRB 20190208A & 5 & 59860.36088288 & $30_{-8}^{+9}$ & $89.6_{-10.4}^{+10.4}$ & $-0.4_{-7.7}^{+7.7}$ & $1000_{-1}^{+1}$ & $1102_{-3}^{+3}$\\ 
\midrule
FRB 20190303A & 1 & 59631.79600438 & $-569_{-3}^{+3}$ & $98.5_{-3.5}^{+3.5}$ & $-1.6_{-2.5}^{+2.5}$ & $1001_{-1}^{+1}$ & $1212_{-6}^{+6}$\\ 
FRB 20190303A & 2 & 59631.79703618 & $-595_{-13}^{+11}$ & $109.7_{-18.7}^{+18.7}$ & $12.7_{-12.7}^{+12.7}$ & $1001_{-1}^{+1}$ & $1096_{-9}^{+9}$\\ 
FRB 20190303A & 3 & 59631.79960369 & $-565_{-2}^{+30}$ & $100.9_{-0.8}^{+0.8}$ & $0.0_{-0.5}^{+0.5}$ & $1001_{-1}^{+1}$ & $1291_{-3}^{+3}$\\ 
FRB 20190303A & 4 & 59631.80357370 & $-556_{-6}^{+20}$ & $103.1_{-3.8}^{+3.8}$ & $0.1_{-2.7}^{+2.7}$ & $1001_{-1}^{+1}$ & $1188_{-2}^{+2}$\\ 
FRB 20190303A & 5 & 59737.58729765 & $-588_{-10}^{+11}$ & $78.9_{-14.7}^{+14.7}$ & $-3.5_{-11.5}^{+11.5}$ & $1189_{-2}^{+2}$ & $1360_{-3}^{+3}$\\ 
FRB 20190303A & 6 & 59792.45869712 & $-832_{-8}^{+8}$ & $136.4_{-18.6}^{+18.6}$ & $-13.2_{-11.1}^{+11.1}$ & $1001_{-1}^{+1}$ & $1148_{-2}^{+2}$\\ 
FRB 20190303A & 7 & 59829.34767827 & $-894_{-6}^{+9}$ & $99.2_{-13.8}^{+13.8}$ & $-7.7_{-9.8}^{+9.8}$ & $1000_{-1}^{+1}$ & $1156_{-2}^{+2}$\\ 
FRB 20190303A & 8 & 59829.35848203 & $-877_{-5}^{+5}$ & $89.1_{-10.3}^{+10.3}$ & $-18.9_{-7.8}^{+7.8}$ & $1024_{-3}^{+3}$ & $1311_{-3}^{+3}$\\ 
FRB 20190303A & 9 & 59829.36416311 & $-917_{-8}^{+8}$ & $107.0_{-16.7}^{+16.7}$ & $13.3_{-11.5}^{+11.5}$ & $1024_{-2}^{+2}$ & $1219_{-5}^{+5}$\\ 
FRB 20190303A & 10 & 59829.36511317 & $-893_{-2}^{+2}$ & $97.3_{-2.2}^{+2.2}$ & $-3.8_{-1.6}^{+1.6}$ & $1025_{-2}^{+2}$ & $1362_{-2}^{+2}$\\ 
FRB 20190303A & 11 & 59846.19271196 & $-751_{-4}^{+4}$ & $85.8_{-7.6}^{+7.6}$ & $-11.0_{-5.8}^{+5.8}$ & $1000_{-1}^{+1}$ & $1250_{-2}^{+2}$\\ 
FRB 20190303A & 12 & 59846.21049448 & - & - & - & $1112_{-3}^{+3}$ & $1396_{-3}^{+3}$\\ 
FRB 20190303A & 13 & 59874.09966469 & $-646_{-8}^{+14}$ & $99.2_{-6.1}^{+6.1}$ & $7.7_{-4.3}^{+4.3}$ & $1000_{-1}^{+1}$ & $1112_{-12}^{+12}$\\ 
\midrule
FRB 20190417A & 1 & 59490.46379059 & $4383_{-5}^{+3}$ & $67.8_{-6.8}^{+6.8}$ & $1.0_{-5.6}^{+5.6}$ & $1001_{-1}^{+1}$ & $1221_{-6}^{+6}$\\ 
FRB 20190417A & 2 & 59490.46897234 & $4454_{-11}^{+14}$ & $76.6_{-5.3}^{+5.3}$ & $-5.4_{-4.2}^{+4.2}$ & $1001_{-1}^{+1}$ & $1140_{-3}^{+3}$\\ 
FRB 20190417A & 3 & 59490.48172190 & - & - & - & $1001_{-1}^{+1}$ & $1235_{-4}^{+4}$\\ 
FRB 20190417A & 4 & 59490.48696321 & $4418_{-8}^{+9}$ & $134.0_{-23.7}^{+23.7}$ & $31.9_{-14.9}^{+14.9}$ & $1195_{-3}^{+3}$ & $1437_{-11}^{+11}$\\ 
FRB 20190417A & 5 & 59490.49411557 & - & - & - & $1001_{-1}^{+1}$ & $1113_{-3}^{+3}$\\ 
FRB 20190417A & 6 & 59490.49483976 & $4365_{-4}^{+4}$ & $85.6_{-6.6}^{+6.6}$ & $-0.1_{-5.0}^{+5.0}$ & $1118_{-2}^{+2}$ & $1424_{-2}^{+2}$\\ 
FRB 20190417A & 7 & 59490.50704840 & - & - & - & $1147_{-1}^{+1}$ & $1374_{-1}^{+1}$\\ 
FRB 20190417A & 8 & 59490.50770092 & - & - & - & $1001_{-1}^{+1}$ & $1125_{-4}^{+4}$\\ 
FRB 20190417A & 9 & 59490.51014115 & - & - & - & $1235_{-2}^{+2}$ & $1499_{-1}^{+1}$\\ 
FRB 20190417A & 10 & 59638.04777479 & - & - & - & $1003_{-5}^{+5}$ & $1249_{-17}^{+17}$\\ 
FRB 20190417A & 11 & 59638.04804896 & $4845_{-14}^{+19}$ & $93.4_{-10.1}^{+10.1}$ & $7.6_{-7.4}^{+7.4}$ & $1406_{-2}^{+2}$ & $1500_{-1}^{+1}$\\ 
FRB 20190417A & 12 & 59638.05282371 & $4733_{-5}^{+7}$ & $72.2_{-6.6}^{+6.6}$ & $6.4_{-5.3}^{+5.3}$ & $1185_{-3}^{+3}$ & $1481_{-9}^{+9}$\\ 
FRB 20190417A & 13 & 59638.05460082 & $4740_{-3}^{+3}$ & $71.2_{-1.2}^{+1.2}$ & $11.9_{-1.0}^{+1.0}$ & $1000_{-1}^{+1}$ & $1289_{-15}^{+15}$\\ 
FRB 20190417A & 14 & 59638.06021707 & $4791_{-12}^{+26}$ & $67.2_{-10.1}^{+10.1}$ & $7.4_{-8.4}^{+8.4}$ & $1137_{-2}^{+2}$ & $1333_{-8}^{+8}$\\ 
FRB 20190417A & 15 & 59638.06269574 & $4622_{-6}^{+5}$ & $54.2_{-7.2}^{+7.2}$ & $1.3_{-6.3}^{+6.3}$ & $1051_{-10}^{+10}$ & $1306_{-2}^{+2}$\\ 
FRB 20190417A & 16 & 59638.06548949 & $4768_{-4}^{+25}$ & $84.9_{-7.5}^{+7.5}$ & $26.2_{-5.9}^{+5.9}$ & $1100_{-3}^{+3}$ & $1382_{-2}^{+2}$\\ 
FRB 20190417A & 17 & 59647.11286895 & $4897_{-7}^{+7}$ & $68.0_{-10.6}^{+10.6}$ & $3.5_{-8.8}^{+8.8}$ & $1069_{-3}^{+3}$ & $1296_{-19}^{+19}$\\ 
FRB 20190417A & 18 & 59650.10934352 & $4850_{-10}^{+8}$ & $78.5_{-12.5}^{+12.5}$ & $-1.0_{-9.8}^{+9.8}$ & $1000_{-1}^{+1}$ & $1162_{-2}^{+2}$\\ 
FRB 20190417A & 19 & 59707.93043242 & - & - & - & $1000_{-1}^{+1}$ & $1160_{-3}^{+3}$\\ 
FRB 20190417A & 20 & 59707.93331452 & $3996_{-8}^{+12}$ & $26.6_{-5.4}^{+5.4}$ & $10.0_{-5.2}^{+5.2}$ & $1009_{-3}^{+3}$ & $1241_{-5}^{+5}$\\ 
FRB 20190417A & 21 & 59707.93576150 & $3955_{-5}^{+7}$ & $71.1_{-7.4}^{+7.4}$ & $0.8_{-6.0}^{+6.0}$ & $1198_{-7}^{+7}$ & $1465_{-7}^{+7}$\\ 
FRB 20190417A & 22 & 59707.93884360 & $3987_{-3}^{+3}$ & $82.9_{-2.4}^{+2.4}$ & $9.2_{-1.9}^{+1.9}$ & $1073_{-3}^{+3}$ & $1351_{-2}^{+2}$\\ 
FRB 20190417A & 23 & 59707.93917490 & $3949_{-14}^{+10}$ & $55.0_{-11.1}^{+11.1}$ & $18.3_{-9.9}^{+9.9}$ & $1127_{-2}^{+2}$ & $1333_{-32}^{+32}$\\ 
FRB 20190417A & 24 & 59707.94032917 & $3946_{-16}^{+14}$ & $67.2_{-10.5}^{+10.5}$ & $13.1_{-8.8}^{+8.8}$ & $1000_{-1}^{+1}$ & $1069_{-2}^{+2}$\\ 
FRB 20190417A & 25 & 59714.89757849 & $4495_{-10}^{+10}$ & $34.8_{-6.0}^{+6.0}$ & $14.1_{-5.8}^{+5.8}$ & $1000_{-1}^{+1}$ & $1141_{-4}^{+4}$\\ 
FRB 20190417A & 26 & 59714.90212106 & $4391_{-11}^{+15}$ & $47.8_{-7.1}^{+7.1}$ & $5.9_{-6.4}^{+6.4}$ & $1000_{-1}^{+1}$ & $1184_{-3}^{+3}$\\ 
FRB 20190417A & 27 & 59714.90596120 & $4507_{-2}^{+4}$ & $96.5_{-7.4}^{+7.4}$ & $1.3_{-5.4}^{+5.4}$ & $1042_{-3}^{+3}$ & $1314_{-2}^{+2}$\\ 
FRB 20190417A & 28 & 59714.90948204 & - & - & - & $1000_{-1}^{+1}$ & $1074_{-6}^{+6}$\\ 
FRB 20190417A & 29 & 59734.82652577 & $4620_{-7}^{+8}$ & $67.4_{-6.5}^{+6.5}$ & $6.2_{-5.4}^{+5.4}$ & $1071_{-3}^{+3}$ & $1290_{-3}^{+3}$\\ 
FRB 20190417A & 30 & 59734.82953117 & $4616_{-11}^{+12}$ & $86.8_{-9.4}^{+9.4}$ & $11.1_{-7.1}^{+7.1}$ & $1000_{-1}^{+1}$ & $1090_{-2}^{+2}$\\ 
FRB 20190417A & 31 & 59743.83752278 & $4487_{-3}^{+4}$ & $72.7_{-4.3}^{+4.3}$ & $12.4_{-3.5}^{+3.5}$ & $1017_{-4}^{+4}$ & $1289_{-3}^{+3}$\\ 
FRB 20190417A & 32 & 59751.75454043 & $4586_{-9}^{+8}$ & $71.3_{-4.2}^{+4.2}$ & $35.7_{-3.7}^{+3.7}$ & $1000_{-1}^{+1}$ & $1136_{-4}^{+4}$\\ 
FRB 20190417A & 33 & 59772.78092918 & $5000_{-5}^{+6}$ & $71.7_{-7.5}^{+7.5}$ & $-0.6_{-6.1}^{+6.1}$ & $1000_{-1}^{+1}$ & $1238_{-47}^{+47}$\\ 
FRB 20190417A & 34 & 59772.78885134 & $5002_{-5}^{+6}$ & $85.0_{-4.3}^{+4.3}$ & $7.3_{-3.3}^{+3.3}$ & $1233_{-3}^{+3}$ & $1500_{-1}^{+1}$\\ 
FRB 20190417A & 35 & 59772.79076582 & $5018_{-11}^{+11}$ & $79.1_{-13.4}^{+13.4}$ & $13.8_{-10.6}^{+10.6}$ & $1080_{-17}^{+17}$ & $1251_{-2}^{+2}$\\ 
FRB 20190417A & 36 & 59772.79140992 & $5042_{-13}^{+13}$ & $91.7_{-6.2}^{+6.2}$ & $15.2_{-4.6}^{+4.6}$ & $1353_{-2}^{+2}$ & $1500_{-1}^{+1}$\\ 
FRB 20190417A & 37 & 59789.61021880 & - & - & - & $1008_{-1}^{+1}$ & $1199_{-4}^{+4}$\\ 
FRB 20190417A & 38 & 59789.61885655 & $5015_{-8}^{+7}$ & $84.3_{-11.3}^{+11.3}$ & $31.2_{-9.0}^{+9.0}$ & $1139_{-3}^{+3}$ & $1374_{-12}^{+12}$\\ 
FRB 20190417A & 39 & 59791.70805074 & $5225_{-21}^{+22}$ & $55.3_{-8.5}^{+8.5}$ & $2.5_{-7.4}^{+7.4}$ & $1000_{-1}^{+1}$ & $1090_{-6}^{+6}$\\ 
FRB 20190417A & 40 & 59791.71641876 & $5096_{-4}^{+5}$ & $73.5_{-8.1}^{+8.1}$ & $-3.7_{-6.5}^{+6.5}$ & $1103_{-2}^{+2}$ & $1332_{-2}^{+2}$\\ 
FRB 20190417A & 41 & 59791.72189893 & $5133_{-6}^{+6}$ & $100.2_{-6.3}^{+6.3}$ & $8.8_{-4.5}^{+4.5}$ & $1233_{-2}^{+2}$ & $1467_{-2}^{+2}$\\ 
FRB 20190417A & 42 & 59791.72525023 & $5140_{-6}^{+10}$ & $81.7_{-9.6}^{+9.6}$ & $14.6_{-7.5}^{+7.5}$ & $1014_{-5}^{+5}$ & $1248_{-2}^{+2}$\\ 
FRB 20190417A & 43 & 59791.72601661 & $5145_{-5}^{+5}$ & $85.7_{-11.6}^{+11.6}$ & $17.2_{-8.9}^{+8.9}$ & $1019_{-5}^{+5}$ & $1238_{-2}^{+2}$\\ 
FRB 20190417A & 44 & 59799.63733950 & $5041_{-10}^{+9}$ & $62.5_{-5.4}^{+5.4}$ & $27.8_{-4.7}^{+4.7}$ & $1187_{-3}^{+3}$ & $1398_{-8}^{+8}$\\ 
FRB 20190417A & 45 & 59799.64065239 & $5016_{-6}^{+9}$ & $61.3_{-10.2}^{+10.2}$ & $14.6_{-8.8}^{+8.8}$ & $1086_{-2}^{+2}$ & $1258_{-3}^{+3}$\\ 
FRB 20190417A & 46 & 59804.67073391 & $4818_{-11}^{+9}$ & $83.4_{-16.1}^{+16.1}$ & $-14.4_{-12.5}^{+12.5}$ & $1000_{-1}^{+1}$ & $1116_{-2}^{+2}$\\ 
FRB 20190417A & 47 & 59804.67273573 & $4848_{-13}^{+8}$ & $81.2_{-13.1}^{+13.1}$ & $9.1_{-10.2}^{+10.2}$ & $1283_{-2}^{+2}$ & $1483_{-3}^{+3}$

\end{longtable}
\endgroup

\begin{multicols}{2}

For FRBs~20190208A and 20190303A, all bursts are highly linearly polarized, with linear polarization fractions $\gtrsim$ 80\%. Sixty-one percent of the bursts have linear polarization fractions greater than 90\%. The median linear polarization fraction is 89.6\% for FRB~20190208A and 99.2\% for FRB~20190303A, both much larger than the median of 63\% observed in the non-repeating FRB sample from Ref.~\cite{2024ApJ...968...50P}. In comparison, the median linear polarization fractions of FRB~20201124A (95.5\%, \cite{jiangraa}), FRB~20220912A (96.0\%, \cite{zhang2023}), and FRB~20240114A (93.6\%, \cite{xie2024FRB20240114A}) are also close to those observed in our sample.
We further conducted a t-test to examine the similarity in linear polarization between repeating and non-repeating FRBs. 
For FRBs~20190208A and 20190303A, we used our FAST results just reported. For FRBs~20201124A (536 bursts), 20220912A (1076 bursts), and 20240114A (299 bursts), we used the samples from Ref.~\cite{jiangraa}, \cite{zhang2023}, and \cite{xie2024FRB20240114A}, respectively. For the non-repeating FRBs, we used the 89 measured samples from Ref.\cite{2024ApJ...968...50P}. For repeaters, all the samples used are detections. To maintain consistency, we did not include upper limits from nondetections for non-repeaters in Ref.\cite{2024ApJ...968...50P}.
The results are shown in Figure~\ref{fig:linear_tt}. In Figure~\ref{fig:linear_tt}, we also show their linear polarization distributions. For all pairwise comparisons among the repeating FRBs, the p-values were greater than 0.05 (corresponding to a logarithmic value of -1.3), suggesting that the linear polarization distributions among the five repeating FRBs are nearly identical. However, the p-values for comparisons between repeating and non-repeating FRBs were significantly smaller than those for comparisons among the repeating FRBs. The box plot, along with the t-test statistics and p-values, clearly reveals a distinct difference in the linear polarization distributions between the non-repeating and the five repeating FRBs.

For FRB~20190417A, some bursts exhibit a low linear polarization fraction. As discussed in Ref.~\cite{feng22}, the depolarization is likely caused by RM scattering due to multi-path transmission of signals in an inhomogeneous magneto-ionic environment. We parameterize the depolarization due to RM scattering \citep{2012MNRAS.421.3300O} as :
\begin{equation}
\label{eq:rmscatter}
f_{\rm{RM~scattering}} \equiv 1 - \exp{(-2\lambda^4\sigma^2_{\mathrm{RM}})} \>\>,
\end{equation}
where \( f_{\rm{RM~scattering}} \) is the fractional reduction in the linear polarization amplitude, \( \sigma_{\mathrm{RM}} \) is the standard deviation of the RM, and \( \lambda \) is the wavelength.

We determine \( \sigma_{\mathrm{RM}} \) by fitting the data with the model in Eq.~(\ref{eq:rmscatter}), assuming each burst has 100\% intrinsic linear polarization. For FRB~20190417A, a weighted least squares fit was used. For \( \lambda \) in Eq.~(\ref{eq:rmscatter}), we calculated the central frequency of each burst and converted the weighted frequency to the equivalent wavelength, \( \lambda \). Additionally, for FRB~20190417A, we included the CHIME measurement with a linear polarization fraction of 14.7\% at 665\,MHz from Ref.~\cite{chime_repeaterRM}. The resulting \( \sigma_{\mathrm{RM}} \) for FRB~20190417A is \( 5.19 \pm 0.09 \,\mathrm{rad\,m^{-2}} \), which is consistent with the measurements of \( 6.1 \pm 0.5 \,\mathrm{rad\,m^{-2}} \) in Ref.~\cite{feng22}.

\noindent
\begin{minipage}{\linewidth}
    \centering
   \includegraphics[width=0.88\linewidth]{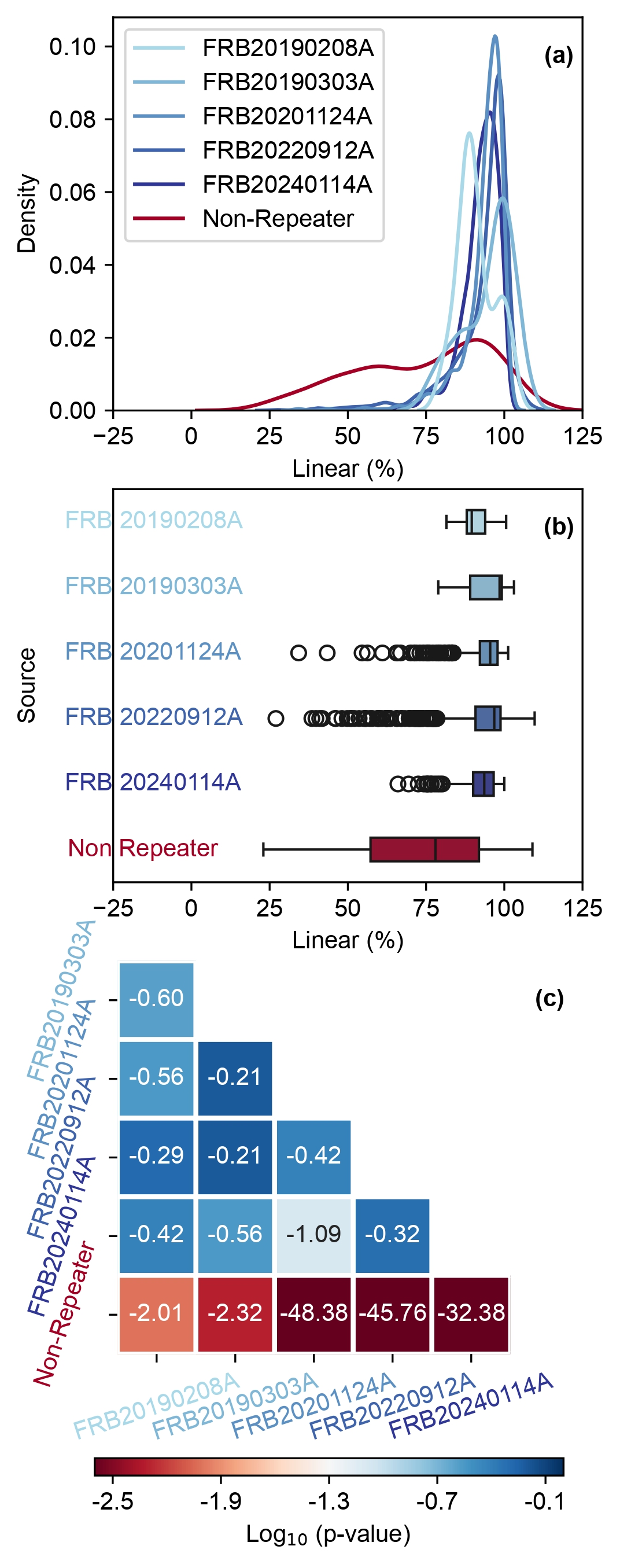}
\captionof{figure}{{\bf Comparison of linear polarization between repeating and non-repeating FRBs.} \textbf{Panel (a)} displays the kernel density estimates (KDE) of the linear polarization degree for five repeating FRBs (blue line) and non-repeating FRBs. \textbf{Panel (b)} presents a boxplot of the linear polarization for the six datasets, where the boxes represent the quartiles of the datasets, and the whiskers extend to the maximum and minimum values after excluding the extreme values. Extreme values are indicated by black circles and are defined as values exceeding 1.5 times the inter-quartile range. \textbf{Panel (c)} shows the logarithm of the p-values from the pairwise t-test comparisons between the six datasets.}
    \label{fig:linear_tt}
\end{minipage}

\begin{center}
\renewcommand{\arraystretch}{1.15}
\setlength{\tabcolsep}{2pt}
\begin{table*}[!t]
\centering
\caption{Rotation measurements of 18 Repeating FRBs. `-' represents no reliable measurement. Column (1): FRB name; Column (2): the minimum RM ($\rm{RM_{min}}$); Column (3): the maximum RM ($\rm{RM_{max}}$); Column (4): Difference between the minimum and maximum (${\bigtriangleup}_{\rm{RM}}$); Column (5): the average RM ($\rm{RM_{mean}}$); Column (6): the standard deviation of the RM ($\sigma_{\rm RM}$); Column (7): Whether RM is reversed is marked with `yes' or `no'. Column (8): Reference.}\label{tab:rm}
    \begin{tabular}{cccccccc|cccp{15cm}}
    \toprule 
        \multicolumn{8}{c|}{{\makecell{Repeating FRBs with multiple RM measurements}}} & \multicolumn{3}{c}{{\makecell{Repeating FRBs with\\ only one RM measurement}}} \\
        \hline
        Source name & $\rm{RM_{min}}$ & $\rm{RM_{max}}$ &  ${\bigtriangleup}_{\rm{RM}}$ & $\rm{RM_{mean}}$ & ${\sigma}_{\rm{RM}}$ & RM reversal& Ref. & Source name & $\rm{RM}$ & Ref. \\
        & $\rm{rad~m^{-2}}$ & $\rm{rad~m^{-2}}$ &  $\rm{rad~m^{-2}}$ & $\rm{rad~m^{-2}}$ & $\rm{rad~m^{-2}}$ &  &  & & $\rm{rad~m^{-2}}$ \\
        \hline
        FRB~20121102A & 30755.0 & 126750.0 & 95995.0   & 85001.5  & $30.9\pm0.4$  & no  & \cite{2022MNRAS.511.6033P,121102rm,2021ApJ...908L..10H}& FRB~20180814A & 700.1 & \cite{chime_repeaterRM} \\
        FRB~20180301A & -237.0  & 563.9    & 800.9     & 101.8    & $6.3\pm0.4$   & yes & \cite{luo2020,2023MNRAS.526.3652K}      & FRB~20190222A & 567.7 & \cite{chime_repeaterRM}\\
        FRB~20180916B & -123.3  & -67.3    & 56.0      & -107.8   & $0.12\pm0.01$ & no  & \cite{2023ApJ...950...12M,2021ApJ...911L...3P,2020ApJ...896L..41C,2022ApJ...932...98S}& FRB~20190604A & -17.8 & \cite{chime_repeaterRM} \\
        FRB~20181030A & 36.5    &  38.5    & 2.0       & 37.2     & -             & no  & \cite{chime_repeaterRM}        & FRB~20190711A & 9.0 & \cite{askap20} \\
        FRB~20181119A & 479.2   & 1339.3   & 860.1     & 809.1    & -             & no  & \cite{chime_repeaterRM}       &  &  & \\
        FRB~20190117A & -165.0  & 79.6     & 244.6     & -4.8     & $2.78\pm0.05$ & yes & \cite{chime_repeaterRM}       &  & & \\
        FRB~20190208A & 10.1    & 32.1     & 22.5      & 22.8     & -             & no  & \cite{chime_repeaterRM}      &  &  & \\
        FRB~20190212B & -5.3    & -0.1     & 5.2       & -1.2     & -             & no  &  \cite{chime_repeaterRM}       &  & & \\
        FRB~20190303A & -917.0  & -205.4   & 711.6     & -594.2   & $3.6\pm0.1$   & no  & \cite{chime_repeaterRM}       &  & & \\
        FRB~20190417A & 3946.0  & 5225.0   & 1279.0    & 4682.6   & $5.19\pm0.09$ & no  &  \cite{chime_repeaterRM}     &  & & \\
        FRB~20190520B & -24034.5& 12956.0  & 36990.5   & -779.2   & $218.9\pm10.2$& yes & \cite{reshma23}   &  & & \\
        FRB~20200120E & -57.2   & -21.9    & 35.3      & -38.2    & -             & no  & \cite{2022NatAs...6..393N}       &  & & \\
        FRB~20201124A & -887.2  & -362.7   & 524.5     & -586.4   & $2.5\pm0.1$   & no  & \cite{2022Natur.609..685X,2021MNRAS.508.5354H,2022MNRAS.512.3400K} &  & & \\
        FRB~20220912A & -12.5   & 20.5     & 33.0      & 0.1      & -             & no\textsuperscript{*} & \cite{2023ApJ...955..142Z} &  & & \\
    \bottomrule
    \end{tabular}
    \begin{tablenotes}
    \begin{itemize}
      \item[\textsuperscript{*}]: FRB~20220912A is excluded because its RM variations are smaller than 50$\,\mathrm{rad\,m^{-2}}$, suggesting that the variations may not be due to the FRB's local environment.
\end{itemize}      
\end{tablenotes}
\end{table*}
\end{center}

\noindent
 \begin{minipage}{\linewidth}
    \centering
   \includegraphics[width=0.9\linewidth]{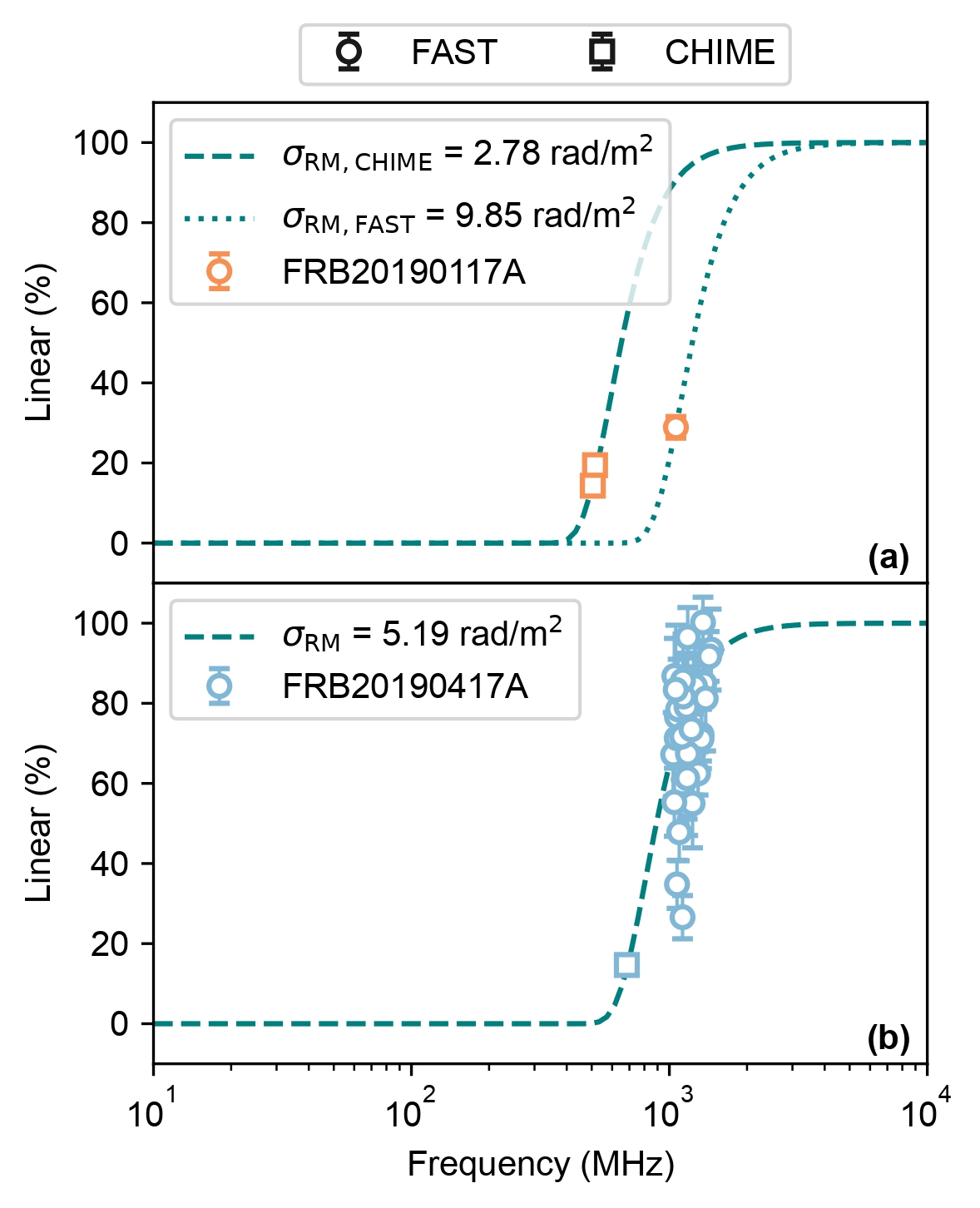}
\captionof{figure}{{\bf Measurements of \( \sigma_{\rm RM} \) for FRB 20190117A and 20190417A.} Square and circle markers indicate CHIME and FAST data points, respectively. In \textbf{panel (a)}, the dashed line and dotted line represent the respective fitting of \( \sigma_{\rm RM} \) for the CHIME and FAST observational data points of FRB~20190117A. In \textbf{panel (b)}, the dashed line represents the fitting of \( \sigma_{\rm RM} \) for the FAST observational data points of FRB~20190417A.}
    \label{fig:sigmarm}
\end{minipage}

For FRB~20190117A, we detected only one burst with a linear polarization fraction of 28.9\% at 1060\,MHz. We also included two CHIME bursts, with linear polarization fractions of 19.5\% at 515\,MHz and 14.3\% at 505\,MHz. The resulting \( \sigma_{\mathrm{RM}} \) for FRB~20190117A is 9.85\,rad\,m\(^{-2}\) for the FAST data and \( 2.78 \pm 0.05 \,\mathrm{rad\,m^{-2}} \) for the CHIME data, respectively. The result of the fitting is shown in Figure~\ref{fig:sigmarm}. We did not use all the data from CHIME and FAST for the fit because the \( \sigma_{\mathrm{RM}} \) values were not consistent. The inconsistency may be caused by variations in the magneto-ionic environment, as indicated by the RM reversal that occurred within about three weeks of FRB 20190117A, as presented below. Another possibility is that intrinsic effects are at play, and the linear polarization fraction measured at FAST is lower than expected. Such intrinsic effects have already been observed in FRB~20201124A \citep{2022Natur.609..685X} and FRB~20220912A \citep{feng2024}.

\begin{figure*}[!htbp]
    \centering
    \includegraphics[width=0.9\textwidth]{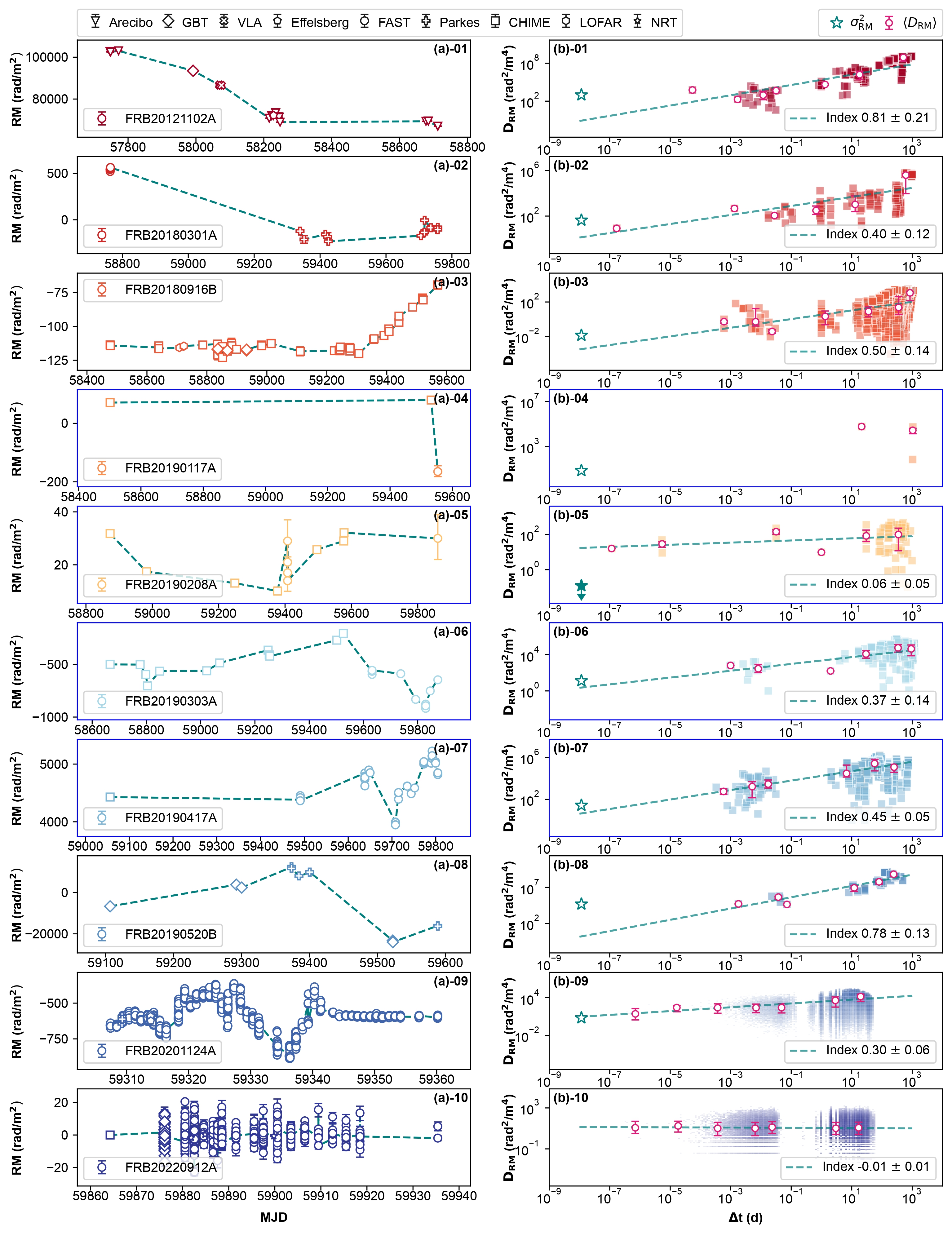}
    \caption{{\bf Variation of RM with time for 10 repeating FRBs and the RM structure function.} \textbf{Panels (a)01-10} show the trend of RM measurements over time for 10 repeating FRBs, with different markers representing measurements from different telescopes. \textbf{Panels (b)01-10} display the corresponding  structure functions of RM variations. The pink dots represent the second-order structure function. 
    The square symbols represent $\left[{\rm RM}(t) - {\rm RM}(t+\Delta t)\right]^2$ for all pairs of observations with a time separation of $\Delta$ t.
    The green dashed line represents a power-law fit to the structure function. The green star symbol indicates the squared value of \( \sigma_{\rm RM} \) at the 1 ms position, while for FRB 20190208A, there is only an upper limit for \( \sigma_{\rm RM} \). The blue spine plots ((a)/(b)04-07) in the figure illustrate the four repeating FRBs described in detail in this paper.}
    \label{fig:rmvary}
\end{figure*}

We show the RM variations of FRBs~20190117A, 20190208A, 20190303A, and 20190417A in the blue spine plots of Figure~\ref{fig:rmvary}. RM measurements from Ref.~\cite{chime_repeaterRM} are also included in the figure.

FRB~20190208A does not show significant RM variations. Our RM measurements are consistent with those from CHIME. Combining the CHIME data, the RM varies between approximately 10 and 32\,rad\,m\(^{-2}\). The small variation is likely due to measurement uncertainties, and FRB~20190208A likely resides in a non-magneto-ionic environment. In contrast, FRBs~20190117A, 20190303A, and 20190417A all show large RM variations: The RM of FRB~20190117A varies between approximately -165 and 80\,rad\,m\(^{-2}\), the RM of FRB~20190303A varies between approximately -917 and -205\,rad\,m\(^{-2}\), and the RM of FRB~20190417A varies between approximately 3946 and 5225\,rad\,m\(^{-2}\). The large RM variations for these three FRBs indicate that they are in a magneto-ionic environment. Notably, FRB~20190117A exhibits an RM reversal, joining the other two repeating FRBs, FRBs~20190520B and 20180301A. We discuss the implications of these RM measurements in the next section.

\section{Discussions} \label{sec:dis}
We summarize the RM characteristics of published repeating FRBs in Table~\ref{tab:rm}. The RM measurements of the four repeating FRBs mentioned in the previous section are listed in the table, along with other published repeating FRBs, for a total of 18 repeating FRBs with RM measurements. Some repeating FRBs have only one RM measurement and are listed in the right column. For repeating FRBs with multiple RM measurements, the table lists the average, minimum, maximum, the difference between the minimum and maximum RM, and \( \sigma_{\rm RM} \). Whether RM is reversed is also indicated with `yes' or `no'.  
For the 14 repeating FRBs with multiple RM measurements, 9 (i.e., 64\%) show RM variations greater than 50\,rad\,m\(^{-2}\). We take 50\,rad\,m\(^{-2}\) as the threshold for RM variations from the FRB's local environment because RM variation amplitudes of 10\,rad\,m\(^{-2}\) to 30\,rad\,m\(^{-2}\) can arise from apparent RM variations across the phase of a burst, intrinsic frequency evolution \cite{2022Natur.609..685X,feng2024}, or large measurement uncertainties. The overall pattern of RM variations in repeating FRBs is consistent with the findings in Refs. \citep{chime_repeaterRM, cherry2025}. We note that RM variations smaller than 50\,rad\,m\(^{-2}\) could result from limited observations or RM measurements. For example, the RM variations of FRB~20180916B and FRB~20190117A have been smaller than 50\,rad\,m\(^{-2}\) over several years. Long-term, high-cadence observations would help reveal the true RM variations. The percentage of 64\% could therefore represent a lower limit for the number of repeating FRBs residing in a dynamic magneto-ionic environment. For the 14 repeating FRBs with multiple RM measurements, 3 (i.e., 21\%) exhibit RM reversal. Similarly, the 21\% figure may also represent a lower limit, as limited observations may not capture the full extent of RM variations.
   


We then discuss how the RM varies by using the structure function of RM variations. Structure function has been used in the analysis of FRB environments \cite{2023ApJ...950...12M,2024arXiv241115546L}. The structure function of RM variations can be described as:
 \noindent
 \begin{minipage}{\linewidth}
    \centering
   \includegraphics[width=0.9\linewidth]{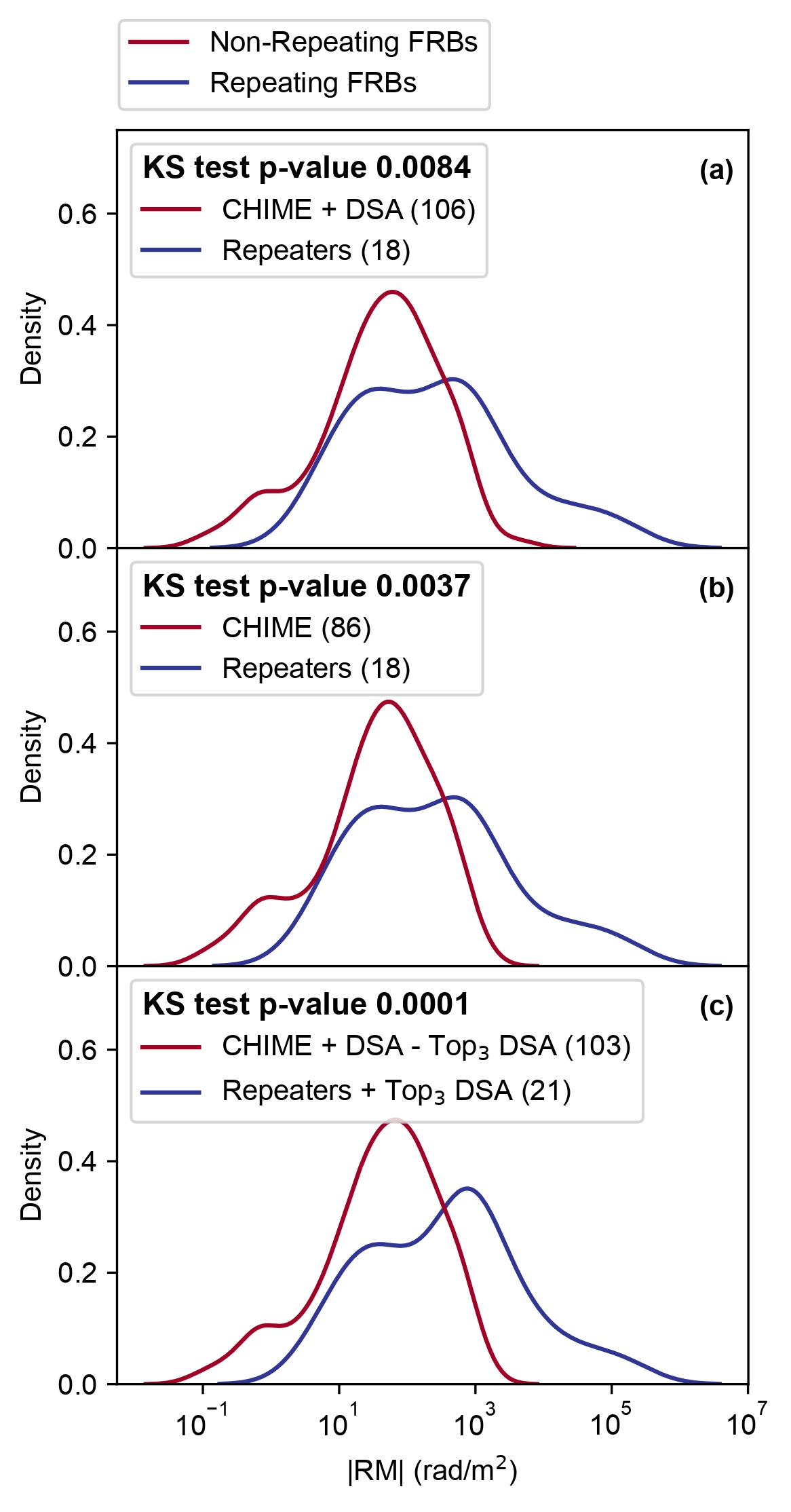}
\captionof{figure}{{\bf Comparison of RM between repeating and non-repeating FRBs.} \textbf{Panel (a)} displays the KDE distribution of RM for 18 repeating FRBs and a total of 106 non-repeating FRBs detected by CHIME and DSA. \textbf{Panel (b)} shows the KDE distribution of RM for the repeating FRBs and 86 non-repeating FRBs detected by CHIME. \textbf{Panel (c)} presents the KDE distribution of RM for repeating and non-repeating FRBs after the three largest RM values from non-repeating FRBs detected by DSA are categorized as part of the repeating FRB sequence. The p-values from the K-S test are 0.0084, 0.0037, and 0.0001, respectively.}
    \label{fig:rm_ks}
\end{minipage}
\[
D_{\rm RM} = \left<\left[{\rm RM}(t) - {\rm RM}(t+\Delta t)\right]^2\right>,
\]
where \( D_{\rm RM} \) represents the structure function of RM variations, corresponding to the average value over a time interval of \( \Delta t \), with the error represented by the quartile range within that \( \Delta t \) interval (indicated by the pink dots in column (b) of Figure~\ref{fig:rmvary}). Subsequently, we fit the structure function of RM variations on a log-log scale using a linear function, i.e., \( \log D_{\rm RM} = \gamma \log \Delta t + C \), where \( \gamma \) corresponds to the power law index. The results are shown in Figure~\ref{fig:rmvary}.

Figure~\ref{fig:rmvary} shows that these FRB repeaters appear to have complex diverse patterns of their RM variations. Most of them exhibit random variation patterns, which might be contributed by a plasma screen with an turbulent medium near the FRB source. Due to the turbulent dynamics or the relative motion between the screen and the FRB source, an irregular RM variation would be caused by the fluctuating clumps along the line of sight, as proposed by Ref.~\cite{2023MNRAS.520.2039Y}. 
In astrophysical plasma, the turbulence is ubiquitous and can naturally induce fluctuations in density and magnetic fields, leading the RM fluctuations.
Assuming that the timescale of the relative motion is shorter than the dynamic timescale of turbulence, the random RM variation would be dominated by the relative motion between the screen and the FRB source. For the uniform relative motion with a transverse velocity as $v_\perp$, the RM structure function $D_{\rm RM}(t)$ in the time domain should be directly related to the RM structure function $D_{\rm RM}(l)$ in space, where $l=v_\perp t$ is the transverse lengthscale on the plasma screen. 
For a given plasma screen, $D_{\rm RM}(l)$  depends on the power spectrum $P(k)$ of the fluctuations of RM density (that is defined by the product of the electron density and the parallel magnetic field), where $k=2\pi/l$ is the spatial wavenumber. In turbulence, the power spectrum usually satisfies a power-law distribution, $P(k)\propto k^{\alpha}$ for $2\pi/L<k<2\pi/l_0$, where $L$ and $l_0$ are outer scale and inner scale, respectively. The power spectrum with $\alpha<-3$ is called a ‘steep spectrum’ (such as Kolmogorov scaling with $\alpha=-11/3$) and the power spectrum with $\alpha>-3$ is called a ‘shallow spectrum’. Fluctuations in the steep and shallow spectra are dominated by large-scale clumps at $\sim L$ and small-scale clumps at $\sim l_0$, respectively.
For a thick plasma screen, the RM structure function satisfies: $D_{\rm RM}\propto l^{-(\alpha+2)}$ for $l<L$ and $D_{\rm RM}\propto {\rm const.}$ for $l>L$, see Ref.~\cite{2023MNRAS.520.2039Y} for a detailed discussion. In particular, Kolmogorov scaling with $\alpha=-11/3$ leads to an RM structure function of $D_{\rm RM}\propto l^{5/3}$ for $l<L$ and  $D_{\rm RM}\propto {\rm const.}$ for $l>L$.
Most FRB repeaters discussed in this work show a power-law index of $\gamma\sim(0-0.8)$, which suggests that the index of the power spectrum of the fluctuation is $\alpha=-(\gamma+2)\sim-(2.0-2.8)$. In this case, the plasma screens near these FRB repeaters have shallow spectra in the inertial ranges (i.e., $l_0<l<L$) due to $\alpha>-3$, if their RM variations are attributed to the turbulence. Therefore, the variation is dominated by small-scale RM density fluctuations. 
The RM density fluctuations depend on the fluctuations of both the electron density and the parallel magnetic field.
Since the magnetic energy spectrum is usually steep as proposed in the literature \citep{Goldreich1995, Lazarian1999, Bowen2024}, the observed results of the shallow spectra imply that a shallow electron density spectrum is more likely to dominate the RM fluctuations.
Physically, a shallow electron density spectrum naturally arises in supersonic turbulence, e.g. in star-forming regions \citep{Hennebelle2012,Xu2017}.


 \noindent
 \begin{minipage}{\linewidth}
    \centering
   \includegraphics[width=0.9\linewidth]{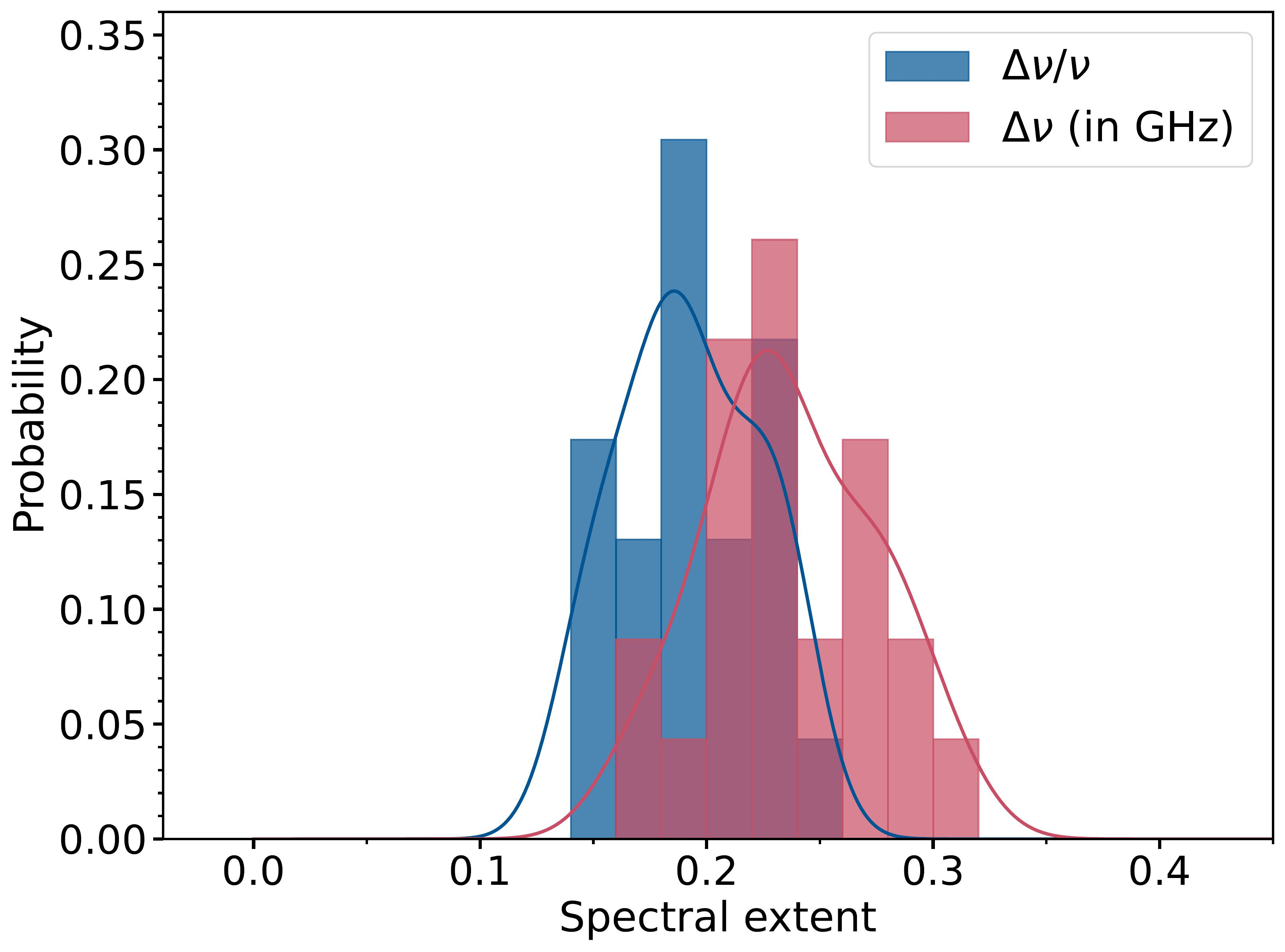}
\captionof{figure}{{\bf Distribution of the spectral extents of the FAST-detected bursts from FRB~20190417A.} Bursts that reached the edges of the FAST frequency band were excluded. The spectral extents are characterized by the fractional bandwidth (\( \Delta \nu / \nu \)) and absolute bandwidth (\(\Delta\nu \)), with their respective distributions shown using Gaussian kernel density estimates.}
    \label{fig:spectral_extent}
\end{minipage}

\begin{center}
\setlength{\tabcolsep}{0pt} 
\renewcommand{\arraystretch}{1.5} 
\setlength{\arrayrulewidth}{0.8pt}
\begin{table*}[!htbp]
\centering
\caption{K-S tests comparing RMs of repeating FRBs and non-repeating FRBs. The columns represent different samples of repeating FRBs, while the rows represent samples of non-repeating FRBs. For the RMs of repeating FRBs, two cases were considered. In case one, we selected the maximum $|$RM$|$ for each FRB. In case two, we selected the earliest RM measurement, in order to mimic the selection of RMs for non-repeating FRBs. For each case, we considered three groups of samples, with the number of samples indicated in parentheses. The ``fiducial'' group consists of the 18 repeating FRBs listed in Table~\ref{tab:rm}. The ``fiducial+3'' group includes the 18 repeating FRBs and the three largest $|$RM$|$ values in the DSA sample \citep{dsa_rm}. The ``selection bias'' group excludes $|$RM$|$ values greater than 10000\,rad\,m$^{-2}$ to assess the potential selection bias due to CHIME's insensitivity to RMs greater than $\sim$5000\,rad\,m$^{-2}$. For the RMs of non-repeating FRBs, five groups of samples were considered. The ``CHIME'' group includes the 86 non-repeating FRBs from Ref.~\cite{2024ApJ...968...50P}. The ``DSA'' group includes the 20 non-repeating FRBs from Ref.~\cite{dsa_rm}. The ``CHIME+DSA'' group is the combination of the ``CHIME'' and ``DSA'' groups. The ``CHIME+DSA-3'' group is the ``CHIME+DSA'' group with the three largest $|$RM$|$ values in the DSA sample excluded. The ``DSA-3'' group is the ``DSA'' group with the three largest $|$RM$|$ values excluded from the DSA sample. P-values smaller than 0.05 are highlighted with red cells, while those equal to or greater than 0.05 are highlighted with blue cells.}\label{tab:rm_ks}
    \begin{tabular}{c|c|c|c|c|c|c}
    \hline
        & \multicolumn{3}{c|}{{\makecell{The maximum $|$RM$|$ \ \ \ \ \ \ \ \ \ }}} & \multicolumn{3}{c}{{\makecell{The earliest RM measurement}}} \\
        \hline
	&Fiducial+3 (21)&Fiducial (18)&Selection bias (16)&Fiducial+3 (21)&Fiducial (18)&Selection bias (17)\\
        \hline
        CHIME+DSA (106) & - & \cellcolor{red!50}\parbox[c][0.3cm][c]{1.9cm}{\centering 0.0084} &\cellcolor{blue!50}\parbox[c][0.3cm][c]{2.8cm}{\centering 0.057} & - & \cellcolor{red!50}\parbox[c][0.3cm][c]{1.9cm}{\centering 0.018} & \cellcolor{red!50}\parbox[c][0.3cm][c]{3.0cm}{\centering 0.043} \\
        \hline
	\parbox[c][0.3cm][c]{3.5cm}{\centering CHIME+DSA-3(103)} &  \cellcolor{red!50}\parbox[c][0.3cm][c]{2.3cm}{\centering 0.0001} & \cellcolor{red!50}\parbox[c][0.3cm][c]{1.9cm}{\centering 0.0041} &\cellcolor{red!50}\parbox[c][0.3cm][c]{2.2cm}{\centering 0.033} & \cellcolor{red!50}\parbox[c][0.3cm][c]{2.2cm}{\centering 0.0004} &\cellcolor{red!50}\parbox[c][0.3cm][c]{1.9cm}{\centering 0.0093} &\cellcolor{red!50}\parbox[c][0.3cm][c]{2.2cm}{\centering 0.025} \\
	\hline
	 CHIME (86) & \cellcolor{red!50}\parbox[c][0.3cm][c]{2.2cm}{\centering 0.0001} &\cellcolor{red!50} \parbox[c][0.3cm][c]{1.9cm}{\centering 0.0037} & \cellcolor{red!50}\parbox[c][0.3cm][c]{2.4cm}{\centering 0.030} &\cellcolor{red!50} \parbox[c][0.3cm][c]{2.2cm}{\centering 0.0003} & \cellcolor{red!50}\parbox[c][0.3cm][c]{1.9cm}{\centering 0.0071} & \cellcolor{red!50}\parbox[c][0.3cm][c]{2.2cm}{\centering 0.020} \\	
	\hline
	 DSA (20) & - & \cellcolor{blue!50}\parbox[c][0.3cm][c]{1.9cm}{\centering 0.49} & \cellcolor{blue!50}\parbox[c][0.3cm][c]{2.4cm}{\centering 0.87} & - & \cellcolor{blue!50}\parbox[c][0.3cm][c]{1.9cm}{\centering 0.75} & \cellcolor{blue!50}\parbox[c][0.3cm][c]{2.2cm}{\centering 0.90} \\		
	 \hline
	 DSA-3 (17) &  \cellcolor{red!50}\parbox[c][0.3cm][c]{2.2cm}{\centering 0.029} & \cellcolor{blue!50}\parbox[c][0.3cm][c]{1.9cm}{\centering 0.10} & \cellcolor{blue!50}\parbox[c][0.3cm][c]{2.2cm}{\centering 0.31} & \cellcolor{blue!50}\parbox[c][0.3cm][c]{2.2cm}{\centering 0.066} & \cellcolor{blue!50}\parbox[c][0.3cm][c]{1.9cm}{\centering 0.22} & \cellcolor{blue!50}\parbox[c][0.3cm][c]{2.2cm}{\centering 0.47} \\		
        \hline
    \end{tabular}
\end{table*}
\end{center}

The non-repeating FRBs have only one RM measurement, or even no RM measurement, and thus RM variations cannot be analyzed. However, we can still compare the RMs of non-repeating and repeating FRBs. Ref.~\cite{feng22} compared the RMs of nine repeating FRBs and eleven non-repeating FRBs, and a K-S test between these two groups yields a p-value of 0.02, indicating a statistically significant difference between them. However, the sample size used in that study was only 21. Refs. \citep{chime_repeaterRM, cherry2025} compare the RMs of repeating and non-repeating FRBs. The samples used are all from the CHIME telescope, which may introduce selection bias, as we will discuss below. Ref.~\cite{2024ApJ...968...50P} included RMs from 86 non-repeating FRBs, and Ref.~\cite{dsa_rm} included RMs from 20 non-repeating FRBs. We perform K-S tests comparing RMs of repeating FRBs and non-repeating FRBs. For the RMs of repeating FRBs, two cases were considered. In case one, we selected the maximum $|$RM$|$ for each FRB. In case two, we selected the earliest RM measurement, in order to mimic the selection of RMs for non-repeating FRBs. For each case, three groups of samples were considered:
\begin{itemize}
\item The ``fiducial'' group consists of the 18 repeating FRBs listed in Table~\ref{tab:rm}.
\item The ``fiducial+3'' group includes the 18 repeating FRBs and the three largest $|$RM$|$ values in the DSA sample \citep{dsa_rm}. Since the DSA observations have a limited time span, non-repeating FRBs in the DSA sample could potentially turn into repeating FRBs. Here we assume that the three non-repeating FRBs with the largest $|$RM$|$ values in the DSA sample are actually repeating FRBs.
\item The ``selection bias'' group excludes $|$RM$|$ values greater than 10000\,rad\,m$^{-2}$ to assess the potential selection bias caused by CHIME's insensitivity to RMs greater than $\sim$5000\,rad\,m$^{-2}$ due to intrachannel depolarization \cite{2024ApJ...968...50P}.
\end{itemize} 
For the RMs of non-repeating FRBs, five groups of samples were considered:
\begin{itemize}
\item The ``CHIME'' group includes the 86 non-repeating FRBs from Ref.~\cite{2024ApJ...968...50P}.
\item The ``DSA'' group includes the 20 non-repeating FRBs from Ref.~\cite{dsa_rm}.
\item The ``CHIME+DSA'' group is the combination of the CHIME'' and ``DSA'' groups.
\item The ``CHIME+DSA-3'' group is the ``CHIME+DSA'' group with the three largest $|$RM$|$ values in the DSA sample excluded. Here we assume that the three non-repeating FRBs with the largest $|$RM$|$ values in the DSA sample are actually repeating FRBs.
\item The ``DSA-3'' group is the ``DSA'' group with the three largest $|$RM$|$ values excluded from the DSA sample.
\end{itemize}

The results are shown in Table~\ref{tab:rm_ks}. Additionally, we plot the RM distributions for three comparisons in Figure~\ref{fig:rm_ks} for case one. For case two, the p-value for comparing the ``CHIME+DSA'' group of non-repeating FRBs with the ``fiducial'' group of repeating FRBs is 0.018, which reveals a marginal dichotomy in the distribution of their RMs. We caution that selection bias could play a role, as the p-values of the ``selection bias'' groups are much larger than those of the ``fiducial'' groups. Excluding two large RMs, the p-values can be about ten times larger. For example, p-value of 0.0084 (smaller than 0.05) becomes 0.057 (greater than 0.05). We note that Ref.~\cite{2024ApJ...968...50P} also compared RMs of repeating FRBs and non-repeating FRBs. In Ref.~\cite{2024ApJ...968...50P}, the sample of repeating FRBs includes 13 sources, excluding two FRBs with large RMs, i.e., FRB~20180301A ($|$RM$|$ $\sim$ 600\,rad\,m$^{-2}$) and FRB~20201124 ($|$RM$|$ $\sim$ 900\,rad\,m$^{-2}$), compared to the ``selection bias'' group. Therefore, Ref.~\cite{2024ApJ...968...50P} found no evidence for a dichotomy in the repeating and non-repeating FRB RM distributions. Besides, the p-values of the ``fiducial+3'' groups are much smaller than those of the ``fiducial'' groups. Therefore, the small sample size of repeating FRBs could easily skew the results. Finally, comparing case one and case two, the selection of the maximum $|$RM$|$ and the earliest RM measurement do influence the results of the K-S tests, but generally not enough to change the trend. For the thirteen K-S tests in each case, only one test changed from greater than 0.05 to smaller than 0.05.

We then discuss the spectral extent
of the bursts using the absolute bandwidth $\Delta \nu$ as well as the fractional bandwidth at the burst central emission frequency $\Delta \nu/\nu$. 
The spectral extent distribution of our FRB~20190417A sample, with bursts that reached the edges of the FAST frequency band removed, is shown in Figure~\ref{fig:spectral_extent}. 
Observationally, the fractional bandwidths $\Delta \nu/\nu$ ranges from 0.14 to 0.24. 
Theoretically, when the radiation is monochromatic in the comoving frame, a generic constraint $\Delta\nu/\nu\gtrsim0.58$ arises due to the high-latitude effect, which poses a generic constraint on the relativistic shock model \citep{KQZ2024}.
Magnetospheric models can intrinsically produce narrow spectra due to the much smaller source angular size \citep{2023ApJ...956...67Y,2024A&A...685A..87W}. Notably, both the bunched coherent inverse Compton scattering mechanism \citep{2022ApJ...925...53Z, Qu&Zhang2024} and the bunched coherent Cherenkov radiation mechanism \citep{lzn2023} offer a remarkable advantage in producing a narrower-band spectrum within the magnetosphere of a magnetar owing to the delta-function nature of the spectral power. We caution that the spectral extent distribution may be subject to selection bias, as bursts reaching the edges of the FAST frequency band were excluded, and the FAST bandwidth is limited to only 500 MHz.

\section{Conclusions} \label{sec:con}
We report multi-year polarization measurements of four repeating FRBs initially discovered by the CHIME telescope: FRBs~20190117A, 20190208A, 20190303A, and 20190417A, and analyzed polarization properties combined with published results. Our main results are the following.
\begin{itemize}

\setlength{\itemsep}{0pt}
 \item[1.] We detected 1, 5, 13, 47 bursts for FRBs~20190117A, 20190208A, 20190303A, and 20190417A, respectively, with a total number of 66. 
\item[2.] Two bursts from FRB 20190417A exhibited significant circular polarization, with a signal-to-noise ratio greater than 7 and a maximum circular polarization fraction of 35.7\%. FRBs~20190208A and 20190303A were highly linearly polarized. FRBs~20190117A and 20190417A showed depolarization due to multi-path propagation, with RM scatter values of $\sigma_{\mathrm{RM}} = 2.78 \pm 0.05 \, \text{rad m}^{-2}$ and $5.19 \pm 0.09 \, \text{rad m}^{-2}$, respectively.
\item[3.] The linear polarization distributions of five repeating FRBs—FRBs~20190208A, 20190303A, 20201124A, 20220912A, and 20240114A—are nearly identical. These distributions differ significantly from those of non-repeating FRBs.
\item[4.] FRBs~20190117A, 20190303A, and 20190417A exhibited substantial RM variations between bursts. Across all repeating FRBs with multiple RM measurements, 64\% showed RM variations exceeding $50 \, \text{rad m}^{-2}$, and 21\% demonstrated RM reversals. A significant proportion of repeating fast radio bursts reside in a dynamic magneto-ionic environment. 
\item[5.] The RM structure function follows a power-law index of $\gamma \sim (0-0.8)$, corresponding to a turbulence power spectrum of $\alpha = -(\gamma + 2) \sim -(2.0-2.8)$. These results suggest that small-scale RM density fluctuations dominate the RM variations.
\item[6.] K-S tests comparing the RMs of repeating and non-repeating FRBs show a slight distinction in their RM distributions. We note that the observed dichotomy could be attributed to the small sample size and potential selection biases. Future studies should aim to increase the sample size from different telescopes and observing conditions of both repeating and non-repeating FRBs.

\end{itemize} 

\subsection*{\begin{center}ACKNOWLEDGMENTS\end{center}}
This work is supported by National Natural Science Foundation of China grant No. 12588202, 12203045, 12233002, 12403100, 12103069, 12403042 by the Leading Innovation and Entrepreneurship Team of Zhejiang Province of China grant No. 2023R01008, and by Key R\&D Program of Zhejiang grant No. 2024SSYS0012. Y.P.Y. is supported by the National Natural Science Foundation of China grant No. 12473047 and the National SKA Program of China (2022SKA0130100). YH acknowledges the support from the Xinjiang Tianchi Program. Di Li is a New Cornerstone investigator. This work made use of the data from FAST (Five-hundred-meter Aperture Spherical radio Telescope) (https://cstr.cn/31116.02.FAST). FAST is a Chinese national mega-science facility, operated by National Astronomical Observatories, Chinese Academy of Sciences.


\InterestConflict{The authors declare that they have no conflict of interest.}


\bibliographystyle{scpma-zycai} 
\bibliography{ms}

\end{multicols}
\end{document}